\documentclass[12pt,letter]{article}
\pdfoutput=1
\usepackage{graphicx, epsfig, color,cite}
\usepackage{amsmath}
\usepackage{amssymb}
\usepackage{float}
\usepackage{hyperref}
\usepackage{subcaption}
\def\eslt{\not\!\!\!{E_T}}
\def\to{\rightarrow}

\def\bi{\begin{itemize}}
\def\ei{\end{itemize}}

\def\tchi{\tilde\chi}

\def\ts{\tilde s}
\def\tb{\tilde b}

\def\tst{\tilde t}

\def\tg{\tilde g}

\def\tq{\tilde q}

\def\tz{\widetilde\chi^0}
\def\alt{\lesssim}

\def\be{\begin{equation}}  
\def\ee{\end{equation}}  
\def\bea{\begin{eqnarray}}  
\def\eea{\end{eqnarray}}

\begin{document}
\begin{titlepage}
\begin{flushright}
OU-HEP-250509
\end{flushright}

\vspace{0.5cm}
\begin{center}
%  {\Large \bf Limits on $R$-parity-violating couplings\\
%    from higgsino pair production in natural supersymmetry}
  {\Large \bf Aspects of the WIMP quality problem \\ and $R$-parity violation\\
 in natural supersymmetry with all axion dark matter}
\vspace{1.2cm} \renewcommand{\thefootnote}{\fnsymbol{footnote}}

{\large Howard Baer$^{1}$\footnote[1]{Email: baer@ou.edu },
Vernon Barger$^2$\footnote[2]{Email: barger@pheno.wisc.edu},
Jessica Bolich$^{1}$\footnote[3]{Email: Jessica.R.Bolich-1@ou.edu},\\
Dibyashree Sengupta$^{3,4}$\footnote[7]{Email: Dibyashree.Sengupta@roma1.infn.it} and
%Xerxes Tata$^5$\footnote[4]{Email: tata@phys.hawaii.edu} and
Kairui Zhang$^1$\footnote[5]{Email: kzhang25@ou.edu}
}\\ 
\vspace{1.2cm} \renewcommand{\thefootnote}{\arabic{footnote}}
{\it 
$^1$Homer L. Dodge Department of Physics and Astronomy,
University of Oklahoma, Norman, OK 73019, USA \\[3pt]
}
{\it 
$^2$Department of Physics,
University of Wisconsin, Madison, WI 53706 USA \\[3pt]
}
{\it
$^3$ INFN, Laboratori Nazionali di Frascati,
Via E. Fermi 54, 00044 Frascati (RM), Italy} \\[3pt]
{\it
  $^4$ INFN, Sezione di Roma, c/o Dipartimento di Fisica, Sapienza Università di Roma, Piazzale Aldo Moro 2, I-00185 Rome, Italy}\\[3pt]
%{\it 
%$^5$Department of Physics and Astronomy,
%University of Hawaii, Honolulu, HI 53706 USA \\[3pt]
%}
%
\end{center}

\vspace{0.5cm}
\begin{abstract}
\noindent
In supersymmetric models where the $\mu$ problem is solved via discrete
$R$-symmetries, then both the global $U(1)_{PQ}$ (Peccei-Quinn,
needed to solve the strong CP problem)  and $R$-parity conservation
(RPC, needed for proton stability) are expected to arise as accidental,
approximate symmetries.
Then in some cases, SUSY dark matter is expected
to be all axions since the relic lightest SUSY particles (LSPs) can decay
away via small $R$-parity violating (RPV) couplings.
We examine several aspects of this {\it all axion} SUSY dark matter scenario.
1. We catalogue the operator suppression which is gained from discrete
$R$-symmetry breaking via four two-extra-field base models.
2. We present exact tree-level LSP decay rates
including mixing and phase space effects and compare to results from
simple, approximate formulae.
3. Natural SUSY models are characterized by light higgsinos with mass
$m\sim 100-350$ GeV so that the dominant sparticle production
cross sections at LHC14 are expected to be higgsino pair production
which occurs at the $10^2-10^4$ fb level.
Assuming nature is natural, the lack of an RPV signal from higgsino pair
production in LHC data translates into rather strong upper bounds on nearly
all trilinear RPV couplings in order to render the SUSY signal (nearly)
invisible.
Thus, in natural SUSY models with light higgsinos, the RPV-couplings must
be small enough that the LSP has a rather high quality of RPC.
\end{abstract}
\end{titlepage}
%\pacs{12.60.-i, 95.35.+d, 14.80.Ly, 11.30.Pb}
%12.60.-i   Models beyond the standard model
%95.35.+d   Dark matter

\section{Introduction}
\label{sec:intro}

Supersymmetry is a highly motivated approach to Beyond-the-Standard-Model (BSM)
physics\cite{Martin:1997ns,Chung:2003fi} in that it stabilizes the weak scale under
quantum corrections with energies far beyond the weak scale.
A supersymmetric version of the Standard Model (SM)
can be constructed by starting with the SM gauge symmetry and matter content,
and then elevating all SM fields to superfields\cite{Baer:2006rs}.
For the Minimal Supersymmetric Standard Model (MSSM), two Higgs doublets are
required for anomaly cancellation and to give mass to all the
quarks and leptons.
The MSSM Lagrangian can be constructed from a minimal choice of K\"ahler
potential and a superpotential given by
\bea
W_{MSSM} &\ni & \mu H_uH_d +\left[ ({\bf f}_u)_{ij}{Q}_i H_u U_j^c +({\bf f}_d)_{ij}
  Q_i H_{d} D_j^c +({\bf f}_e)_{ij} L_i H_{d}E_j^c
  +({\bf f}_\nu)_{ij} L_i H_u N_j^c\right]+\frac{1}{2}M_{N_i}N_i^cN_i^c\nonumber\\
&+& W_{RPV}
\label{eq:Wmssm}
\eea
where the $i,j$ are generation indices.
The term $W_{RPV}$ allows for (possibly large) baryon B- and lepton L-
violating processes via the superpotential terms
\be
W_{RPV}\ni \mu_i^\prime L_i H_u + \lambda_{ijk} L_iL_jE_k^c+
\lambda_{ijk}^\prime L_iQ_jD_k^c +\lambda_{ijk}^{\prime\prime}U_i^cD_j^cD_k^c .
\label{eq:Wrpv}
\ee
Of additional concern are the possible non-renormalizable operators
\be
W_5\ni \kappa_{ijkl}^{(1)}Q_iQ_jQ_kL_l/m_P +\kappa_{ijkl}^{(2)}U_i^cU_j^cD_k^cE_l^C/m_P
\label{eq:W5}
\ee
which lead to dimension-5 proton decay operators.
Supersymmetry breaking is accounted for by adding all allowed soft SUSY
breaking terms to the Lagrangian.

There are actually several puzzling problems which emerge when
constructing the MSSM.
\begin{itemize}
\item One problem with Eq. \ref{eq:Wmssm} is that a SUSY conserving mass
parameter $\mu$ appears, which would be expected to be of order $m_P$, but
phenomenologically is required to be of order the soft breaking terms
$m_{soft}$.
This is the infamous $\mu$ problem of the MSSM.\footnote{Twenty solutions
  to the SUSY $\mu$ problem are reviewed in Ref. \cite{Bae:2019dgg}.}
Its solution requires two parts:
1. the $\mu$ term must be suppressed, perhaps due to some new
symmetry requirements and 2. the $\mu$ term is regenerated at or around
the weak scale by $\hat{H}_u\hat{H}_d$ coupling to additional
Beyond-the-MSSM superfields (that may or may not live in the hidden sector),
and where the new fields gain VEVs thus leading to the $\mu$ term.
This can take place either in the K\"ahler potential
(GM\cite{Giudice:1988yz}) or the superpotential (KN\cite{Kim:1983dt}).
\item The gauge invariant terms in Eq. \ref{eq:Wrpv}
  must be either forbidden or else suppressed at a very high level.
The most severe limits on $R$-parity violating (RPV)
couplings are due to proton decay where the measured limits on the
proton lifetime translate to bounds on the products of couplings
\be
\lambda_{11k}^\prime\lambda_{11k}^{\prime\prime}\alt 10^{-25}
\label{eq:pdecay}
\ee
for $m_{\tq}\sim 1$ TeV.
Other strong limits from $n-\bar{n}$ oscillation and double nucleon
decay provide bounds of order $\lambda^{\prime\prime}\alt 10^{-3}-10^{-4}$\cite{Goity:1994dq} (depending on sfermion mass).
In addition, there are strong limits on the bilinear RPV term in
Eq.~\ref{eq:Wrpv} from limits on neutrino masses~\cite{Barbier:2004ez}:
\be
\sqrt{(\sum_i \mu_i^2)/ \mu^2}\alt 3\times 10^{-6}\sqrt{1+\tan^2\beta}.
\label{eq:mu_i}
\ee
For $\mu\sim 200$ GeV (from naturalness) and $\tan\beta =10$, this implies
$\mu_i\alt 6$ MeV.
For more RPV coupling bounds from other processes,
see {\it e.g.} Ref's~\cite{Barger:1989rk,Dreiner:1997uz,Bhattacharyya:1997vv,Allanach:1999ic}.
\item The MSSM is expected to be the low energy effective field theory
  (LE-EFT) which
  arises from some more ultraviolet (UV) complete theory such as string theory.
  In this case, higher dimensional operators should also be present.
The $\kappa_{ijkl}^{(1,2)}$ coefficients are required to be $\alt 10^{-7}$
from experimental limits on $p$-decay\cite{Hinchliffe:1992ad}.
Dimension-5 operators (Eq. \ref{eq:W5}) contributing to proton decay of the form
  $W\ni QQQQ/m_P$ (where $Q$ stands for visible sector left-chiral superfields)
  must also be suppressed or forbidden.

\end{itemize}

Whatever suppresses the above terms should also allow for the MSSM
Yukawa terms in Eq.~\ref{eq:Wmssm} and should allow for neutrino
mass generation.
Thus, to be phenomenologically viable,
the MSSM ought to be supplemented by one or more additional
symmetries which allow for the necessary terms in the superpotential
while suppressing or forbidding the problematic terms.

Any new symmetry used to suppress/allow the above terms should
be consistent with gravitational effects, which means it should
be\cite{Krauss:1988zc} either a gauge symmetry, a discrete gauge
symmetry
or an $R$-symmetry or discrete $R$-symmetry\cite{Ibanez:1991pr}.
As such, the required extra symmetry must be
anomaly-free\cite{Ibanez:1991hv},
at least up to a universal (Green-Schwartz) contribution.
And in order to accommodate gauge coupling unification, it is best if the
new symmetry is consistent with $SU(5)$ or $SO(10)$ grand unified charge
assignments\footnote{Here, we have in mind {\it local} GUT theories
  which are manifested at certain regions of the compactified space from
  string theory\cite{Buchmuller:2005sh,Nilles:2009yd}.}

Usually, a rather ad-hoc imposition of $R$-parity conservation
(RPC)\cite{Barger:1989rk,Dreiner:1997uz,Bhattacharyya:1997vv}
($R=(-1)^{3(B-L)-2s}$ where $s$ is spin) is assumed to avoid the various
strong limits from different $B$- and $L$- violating processes.
While RPC forbids the terms in $W_{RPV}$, it allows the terms in $W_5$.
The terms in $W_5$ can be forbidden by imposing in addition
baryon triality $B_3$\cite{Ibanez:1991hv}.
The combination of RPV and $B_3$ leads to proton hexality $P_6$\cite{Dreiner:2005rd}.
However, as reviewed in Ref. \cite{Chen:2012tia}, the $P_6$ symmetry--
while encouraging-- does not act to forbid the
$\mu$ parameter and the charges do not respect $SU(5)$ or $SO(10)$
GUT symmetries.

In Lee {\it et al.} Ref. \cite{Lee:2011dya}, it is shown that only
discrete $R$-symmetries can forbid the $\mu$ term while remaining
anomaly-free (via a universal Green-Schwartz term)
and maintaining consistency with $SU(5)$ or $SO(10)$ GUT symmetries.
The discrete $R$ symmetries $\mathbb{Z}_n^R$ all have order $n$ as even divisors
of 24: $\mathbb{Z}_4^R$, $\mathbb{Z}_6^R$, $\mathbb{Z}_8^R$,
$\mathbb{Z}_{12}^R$ or $\mathbb{Z}_{24}^R$.
The various $\mathbb{Z}_n^R$ symmetries can arise as remnants of 10-d
Lorentz symmetry of a string manifold
which then gets compactified to 4-dimensions\cite{Nilles:2012cy}.
$R$-symmetries are characterized by the fact that that the superspace $\theta$
co-ordinates carry $R$-charge $+1$ so that the superpotential carries
$R$-charge $+2$.
The discrete $\mathbb{Z}_n^R$ symmetries also forbid the terms in Eqs.~\ref{eq:Wrpv} and \ref{eq:W5}, while allowing the usual Yukawa
and neutrino mass terms.
The $R$-charges of various MSSM superfields under the various
$\mathbb{Z}_n^R$ discrete symmetries are shown in Table~\ref{tab:R},
taken from Lee {\it et al.}.
\begin{table}[!htb]
\renewcommand{\arraystretch}{1.2}
\begin{center}
\begin{tabular}{c|ccccc}
multiplet & ${\bf Z}_{4}^R$ & ${\bf Z}_{6}^R$ & ${\bf Z}_{8}^R$ & ${\bf Z}_{12}^R$ & ${\bf Z}_{24}^R$ \\
\hline
$H_u$ & 0  & 4  & 0 & 4 & 16 \\
$H_d$ & 0  & 0  & 4 & 0 & 12 \\
$Q$   & 1  & 5 & 1 & 5  & 5 \\
$U^c$ & 1  & 5 & 1 & 5  & 5 \\
$E^c$ & 1  & 5 & 1 & 5  & 5 \\
$L$   & 1  & 3 & 5 & 9  & 9 \\
$D^c$ & 1  & 3 & 5 & 9  & 9 \\
$N^c$ & 1  & 1 & 5 & 1  & 1 \\
\hline
\end{tabular}
\caption{Derived MSSM field $R$ charge assignments for various anomaly-free 
discrete ${\bf Z}_{N}^R$ symmetries which are consistent with $SU(5)$ or 
$SO(10)$ unification (from Lee {\it et al.} Ref.~\cite{Lee:2011dya}).
}
\label{tab:R}
\end{center}
\end{table}

Once the $\mu$ term is forbidden, it next has to be regenerated at or around
the weak scale.
We here adopt the Kim-Nilles method\cite{Kim:1983dt} where $\mu$ arises
from the presence of non-renormalizable operators in the
superpotential of the form
\be
W_{KN}\ni \lambda_\mu S_iS_j H_uH_d/m_P .
\ee
If the hidden fields $S_i$ gain VEVs $\langle S_i\rangle\sim \sqrt{m_{weak} m_P}$,
then a $\mu$ term can be generated
\be
\mu\sim \lambda_\mu v_{S_i}^2/m_P \sim m_{weak} .
\ee
For simplicity, we will here restrict ourselves to models with two additional
(PQ sector) fields labeled $X$ and $Y$.

The possible two-extra-field
models are ordered as ``base models'' $B_I$, $B_{II}$, $B_{III}$ and $B_{IV}$
by Bhattiprolu \& Martin\cite{Bhattiprolu:2021rrj} and the superpotential for each is listed
in Table \ref{tab:base}. The first term in the superpotential leads to
a regenerated $\mu$ term once the $X$ and $Y$ fields develop VEVs.
The second term leads to dimension-6 terms which stabilize the
PQ portion of the scalar potential. Base model $B_I$ was implemented by
Murayama et al.\cite{Murayama:1992dj} in a model where neutrino masses were at first forbidden but then generated radiatively. We adopt a hybrid model of this sort
which includes the actual see-saw neutrino terms (hyMSY)\cite{Baer:2018avn}.
The model $B_{II}$ was written by Chun, Choi and Kim\cite{Choi:1996vz} but as with MSY, we
use a hybrid version of this model labeled hyCCK which includes
see-saw neutrino terms. Base model $B_{III}$ was written by Martin\cite{Martin:2000eq}
(hySPM) while base model $B_{IV}$ comes from
Martin\cite{Martin:2000eq} and Babu, Gogoladze and Wang\cite{Babu:2002ic}
(MBGW).
\begin{table}[!htb]
\renewcommand{\arraystretch}{1.2}
\begin{center}
\begin{tabular}{c|cc}
base model & superpotential & PQ(X,Y) \\
\hline
$B_I$ (hyMSY) & $XYH_uH_d+X^3Y$ & $(-1,3)$ \\
$B_{II}$ (hyCCK/GSPQ) & $X^2H_uH_d+X^3Y$ & $(1,-3)$ \\
$B_{III}$ (hySPM) & $Y^2H_uH_d+X^3Y$ & $(-1/3,1)$ \\
$B_{IV}$ (MBGW) & $X^2H_uH_d+X^2Y^2$ & $(1,-1)$ \\
\hline
\end{tabular}
\caption{Four base models\cite{Bhattiprolu:2021rrj} along with
  associated PQ-sector superpotentials and PQ charges of $X$ and $Y$ fields.
}
\label{tab:base}
\end{center}
\end{table}

The four base models are found to obey an (accidental) global $U(1)_{PQ}$
symmetry with PQ charges for the MSSM fields as listed in Table \ref{tab:PQ}
and the $X$ and $Y$ fields as in Table \ref{tab:base} where the PQ charges
are normalized such that $PQ(H_uH_d)=-2$.
By writing the $X$ and $Y$ fields as a radial-times-angular fields,
then the radial fields may develop VEVs $v_x$ and $v_y$ whilst a
combination of the angular fields contains the axion needed for solving
the strong CP problem.
\begin{table}[!htb]
\renewcommand{\arraystretch}{1.2}
\begin{center}
\begin{tabular}{c|c}
multiplet & PQ charge \\
\hline
$Q$   & $q$ \\
$L$   & $2\cos^2\beta$ \\
$U^c$ & $2\cos^2\beta -q$ \\
$D^c$ & $2\sin^2\beta -q$ \\
$E^c$ & $2(\sin^2\beta -\cos^2\beta)$ \\
$N^c$ & 0 \\
$H_u$ & $-2\cos^2\beta$ \\
$H_d$ & $-2\sin^2\beta$ \\
\hline
\end{tabular}
\caption{Derived PQ charge assignments for various MSSM fields.
  $q$ is arbitrary and $\tan\beta \equiv v_u/v_d$.
}
\label{tab:PQ}
\end{center}
\end{table}

In Ref's \cite{Lee:2010gv,Baer:2018avn} and \cite{Bhattiprolu:2021rrj},
it was found that a solution to the axion quality problem could be obtained
from various of the higher $\mathbb{Z}_n^R$ symmetries.
Minimization of the PQ scalar potential-- augmented by soft SUSY breaking terms-- was found to break the $\mathbb{Z}_n^R$ and PQ symmetries leading to the axionic solution to the strong CP problem with $v_{PQ}\sim \sqrt{m_{soft}m_P}\sim 10^{11}$ GeV, in the cosmological sweet spot\cite{Baer:2018avn} as envisioned
by Kim-Nilles\cite{Kim:1983dt}.
With the RPV-terms forbidden by the $\mathbb{Z}_n^R$
symmetry, then it was expected that dark matter would consist of a
WIMP plus SUSY-DFSZ axion\cite{Bae:2013bva,Bae:2013hma,Bae:2014rfa}
admixture. By including the MSSM higgsinos in the the $a\gamma\gamma$
coupling, then it was found that the $a\gamma\gamma$ coupling was highly
suppressed relative to non-SUSY axion models\cite{Bae:2017hlp}.

In Ref. \cite{Baer:2025oid}, the starting point was again to invoke a
$\mathbb{Z}_n^R$ discrete $R$-symmetry as a fundamental symmetry in the context
of a two-extra-field model which allowed for a KN solution to the
SUSY $\mu$ problem and which then displayed an accidental,
approximate global PQ needed to solve the strong CP problem.
In this case, the renormalizable RPV terms were forbidden so one might expect
exact RPC. However, it was noted that Planck-suppressed higher dimensional
operators of the form
\be
W\ni X^mY^n L_iH_u/m_P^{m+n-1},\ \ \ X^pY^q QQQ/m_P^{p+q},\ \ \ X^rY^sQQQQ/m_P^{r+s+1}
\ee
(where $Q$ denotes a generic matter superfield) could now occur.
When the $X$ and $Y$ fields obtain intermediate scale VEVs,
then RPC also emerged as an {\it approximate, accidental discrete symmetry}
of the MSSM.
The trilinear RPV (tRPV) couplings $\lambda$, $\lambda^\prime$ and
$\lambda^{\prime\prime}$ ended up as allowed, but suppressed by
powers of $f_a/m_P$.
For the cases with $\lambda\sim (f_a/m_P)^3\sim 10^{-21}$, then the lightest
SUSY particle-- taken to be the lightest neutral higgsino in natural SUSY models
with low electroweak finetuning measure $\Delta_{EW}$\cite{Baer:2012up}--
would be stable on timescales of order the age of the universe
$\tau_u\sim 4.3\times 10^{17}$ s.
However, in other cases the tRPV couplings could be suppressed at the
level $\lambda\sim f_a/m_P\sim 10^{-7}$.
In these cases, the LSP could be produced thermally in the early universe
but decay before the onset of Big Bang Nucleosynthesis (BBN).
The SUSY DFSZ axion would still be produced via its usual vacuum misalignment
mechanism\cite{Dine:1982ah,Abbott:1982af,Preskill:1982cy}, and so the dark matter arising from SUSY would be all axions
and no WIMPs.
This prediction seems in accord with recent results from the
LZ experiment\cite{LZ:2024zvo} which requires spin-independent
WIMP-proton cross section
\be
\sigma^{SI}(\tchi p )\alt 2\times 10^{-48}\ {\rm cm}^2\ \ \ (LZ2024)
\ee
for a 100 GeV WIMP.
The LZ result puts significant stress even on
natural SUSY models with mainly axion dark matter where the WIMP thermal
relic density is suppressed from the measured dark matter abundance by a
factor $\xi\sim 10-20$\cite{Baer:2013vpa,Baer:2016ucr}.
In such models, the old adage that SUSY predicts a dark matter candidate
is still true: however, the dark matter particle is a SUSY DFSZ axion
instead of the venerated
SUSY WIMP\cite{Goldberg:1983nd,Ellis:1983ew,Jungman:1995df}.
Even so, it is noted in Ref. \cite{Baer:2025oid}, that an
additional suppression of tRPV operators is still needed to be in accord
with the very severe constraints on products of tRPV couplings from
proton decay: $\lambda^\prime_{11k}\lambda^{\prime\prime}_{11k}\alt 10^{-25}$
for $m_{\tq}\sim 1$ TeV.

In the present paper we explore several different avenues emerging from this
change of paradigm. In Sec. \ref{sec:basemodels}, we present more
comprehensive results on the various bilinear, trilinear and dim-5 proton
decay operator suppression from four two-extra-field
base models\cite{Bhattiprolu:2021rrj}
under each of the possible discrete $R$-symmetries found by
Lee et al.\cite{Lee:2010gv}.
Furthermore, in Ref. \cite{Baer:2025oid} an approximate,
order-of-magnitude expression
was used for the LSP decay rate under tRPV. 
In Sec. \ref{sec:LSPdecay}, we evaluate exactly the tree-level LSP decay
rate including all mixing and phase space effects.
These results show that the approximate results from Ref. \cite{Baer:2025oid}
were sufficient for their conclusions.
In Sec. \ref{sec:lhc}, we explore the situation of light higgsino pair
production at LHC in the context of natural SUSY where production
cross sections are of order $10^2-10^4$ fb,
corresponding to $\sim 10^4-10^6$ higgsino pair events in the
LHC Run 2 data sample.
For the RPC case, the higgsino pair
production signals are very difficult to extract from SM background due to the
small visible energy release from the heavier higgsino decays to the
lightest higgsino: the bulk of energy is carried off by the LSP
rest mass\cite{Baer:2011ec}. In the case of tRPV couplings, then the LSP
can decay into 162 different final states. For the $\lambda_{ijk}$
couplings, hard isolated multileptons $+MET$ should be easily seen above
SM backgrounds.
For the $\lambda^{\prime\prime}_{ijk}$ couplings, each LSP will decay to
3-quark final states which should reconstruct resonantly the LSP invariant mass.
For the $\lambda^\prime_{ijk}$ couplings, a mix of hard isolated lepton plus jet
$+MET $ events should be available. 
If the LSP decays are prompt, then LHC experiments should have seen
such events. If the decays are delayed, but within the LHC detectors, then there would be extraordinary jet/lepton bursts at different locales within the detector volume. 
The fact that no such signals have been seen translates into
(model-dependent) limits on the tRPV couplings in natural SUSY models
so as to keep the higgsino  pair events hidden by have the LSP decay far
outside the detector. A summary and conclusions are contained
in Sec. \ref{sec:conclude}.

\section{Operator suppression in four SUSY base models for different
  $\mathbb{Z}_n^R$ discrete $R$-symmetries}
\label{sec:basemodels}

\subsection{Minimization of PQ scalar potential in base models}
\label{ssec:Vmin}

Along with forbidding the MSSM $\mu$ term, the $\mathbb{Z}_n^R$
symmetries also forbid both the terms of Eq. \ref{eq:Wrpv}
and the dimension-5 proton decay operators thus (presumably)
saving the day for proton-decay constraints.
At the same time, the desired Yukawa and see-saw neutrino terms are
allowed. The $\mathbb{Z}_4^R$ $R$-charges are consistent with $SO(10)$
unification\cite{Lee:2010gv} whilst the other $\mathbb{Z}_n^R$ symmetries
are consistent with $SU(5)$.
In fact, the resulting MSSM Lagrangian-- along with the $X$ and $Y$ fields
and after implementation of the  $\mathbb{Z}_n^R$ symmetry--
exhibits an accidental global $U(1)_{PQ}$ symmetry with PQ charges as
in Tables \ref{tab:PQ} and \ref{tab:base}.

\begin{figure}[htb!]
\centering
    {\includegraphics[height=0.4\textheight]{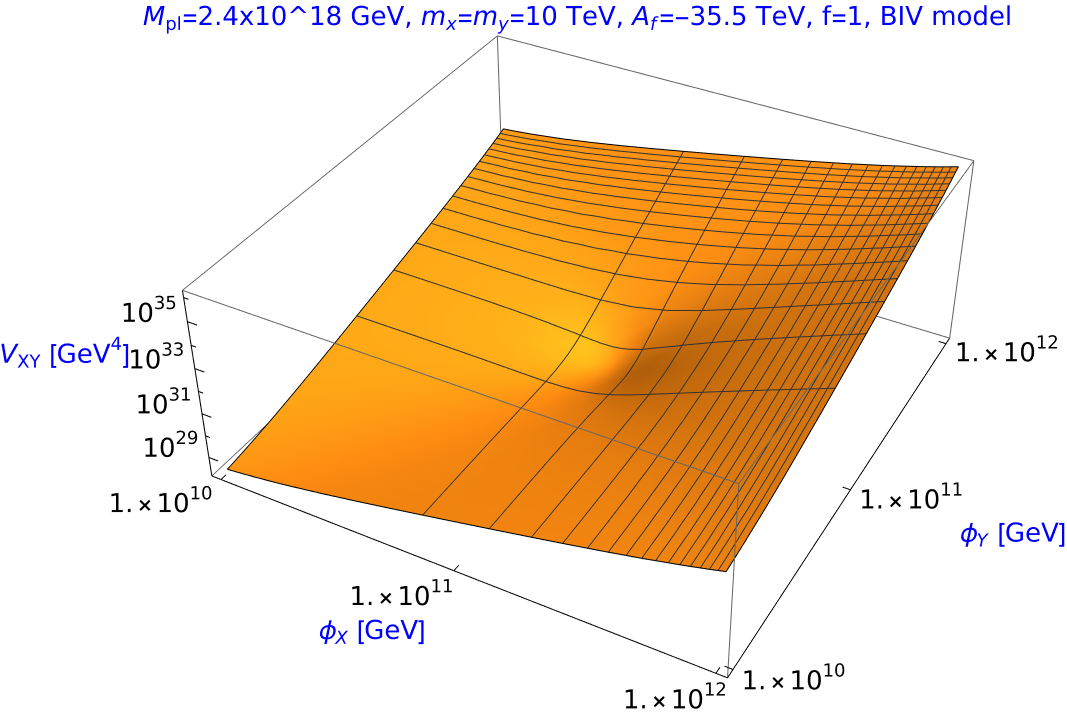}
      \caption{Scalar potential in $\phi_X$ vs. $\phi_Y$ space
        along with minimum for the parameters $m_X=m_Y=10$ TeV
        and $A_f=-35.5$ TeV with $f=1$ for
        base model $B_{IV}$.
        }
     \label{fig:phiXY}}
\end{figure}

From the base model superpotential, we can compute the
corresponding $F$-term of the scalar potential  and augment this with
the appropriate soft SUSY breaking terms. This is done for the base models
$B_I-B_{III}$ in {\it e.g.} Ref's \cite{Bae:2014yta} and \cite{Baer:2018avn}.
The calculation for base model $B_{IV}$ is very similar, but we present
results here for completeness.
From the $B_{IV}$ superpotential listed in Table \ref{tab:base},
$W\ni fX^2Y^2/m_P$, we compute
\be
V_{IV}\equiv\sum_i\left|\frac{\partial W}{\partial\hat{\phi}_i}\right|^2=
4\left[ |f\phi_X\phi_Y^2|^2+|f\phi_X^2\phi_Y|^2\right]/m_P^2
\ee
and augment with
\be
V_{soft}=m_X^2|\phi_X|^2+m_Y^2|\phi_Y|^2+\left(\frac{f A_f\phi_X^2\phi_Y^2}{m_P}+h.c.\right).
\ee
The ensuing scalar potential is plotted in Fig.~\ref{fig:phiXY}
for the case of $m_X=m_Y=10$ TeV, $f=1$ and $A_f=-35.5$ TeV.
The scalar potential admits a minimum at the intermediate scale
$\phi_X=\phi_Y\simeq 10^{11}$ GeV.\footnote{It is possible that with the
  introduction of gauge singlet $X$ and $Y$ fields we introduce the so-called
  destabilizing divergences of Bagger and Poppitz\cite{Bagger:1993ji}.
  Here, we merely assume for now that
  such destabilizing divergences are not a problem.}

\begin{figure}[htb!]
\centering
    {\includegraphics[height=0.4\textheight]{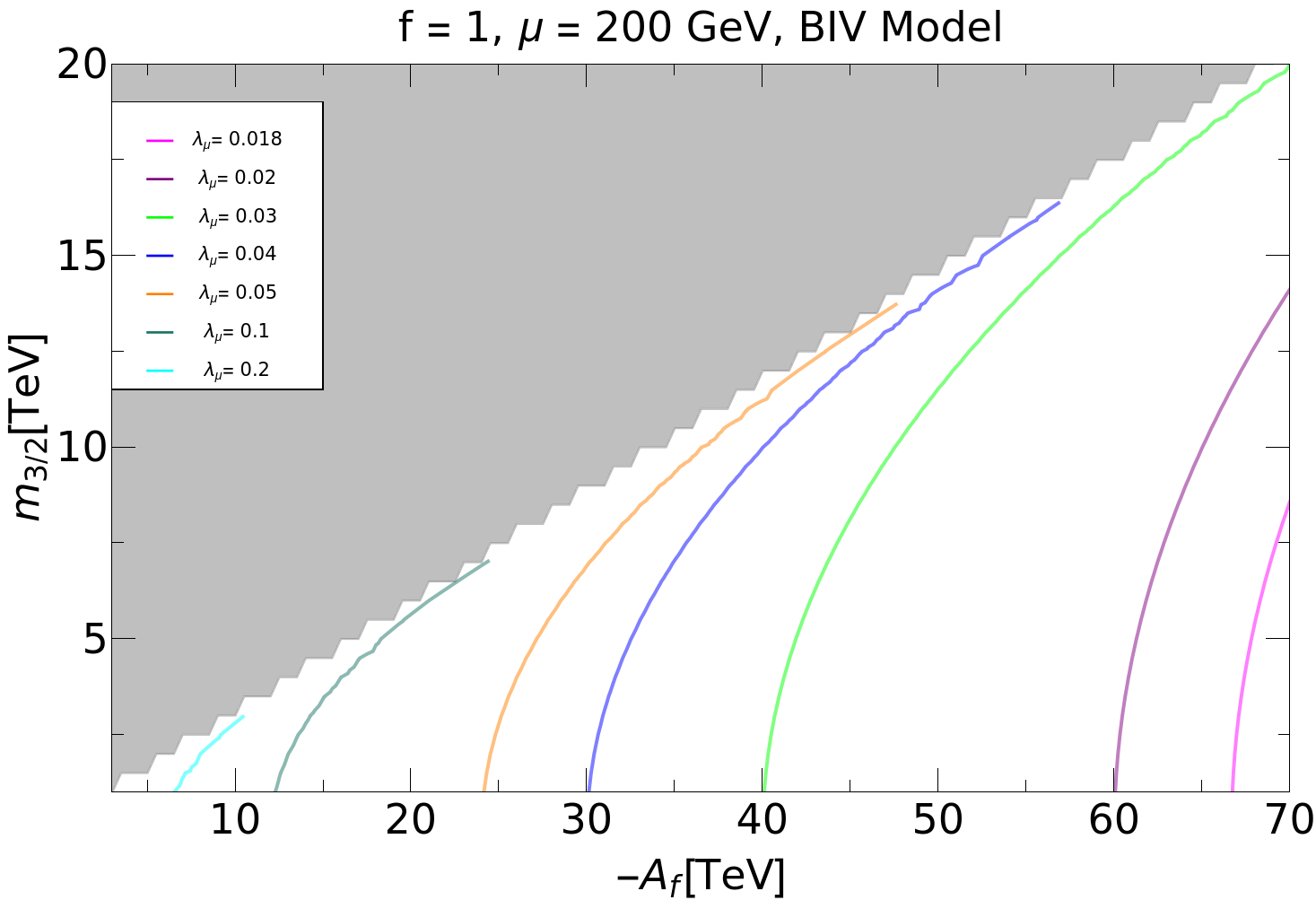}
      \caption{Allowed parameter space for base model $B_{IV}$
        in the $-A_f$ vs. $m_{3/2}$ plane for $\mu =200$ GeV with
        $f=1$ and $\lambda_\mu$ as listed.
        }
     \label{fig:BIVpspace}}
\end{figure}

The MSSM $\mu$ term is then generated with 
\be
W_\mu\ni \lambda_\mu X^2H_uH_d/m_P\ \ \ \mu\sim \lambda_\mu v_X^2/m_P  .
\ee
Contours of $\lambda_\mu$ values leading to a natural value of $\mu =200$ GeV
are then displayed in Fig. \ref{fig:BIVpspace} in the $-A_f$ vs.
$m_{3/2}$ parameter plane, where we take $m_X=m_Y\equiv m_{3/2}$.
As is well-known\cite{Baer:2018avn},
the scalar potential admits spontaneous $\mathbb{Z}_n^R$ and
PQ breaking for large enough values of $-A_f$.

\subsection{R-charge assignments for four base models under five $\mathbb{Z}_n^R$ discrete $R$-symmetries leading to a weak scale value of $\mu$}
\label{ssec:RXY}

Next, we tabulate the various $R$-charge possibilities for the four base models
under the five different $\mathbb{Z}_n^R$ discrete $R$-symmetries.

For base model $B_I$ under $\mathbb{Z}_4^R$, we take $R$-charges of $R(X)=p$
and $R(Y)=q$. Then $p$ and $q$ satisfy:
$p+q = 2+4n$ and $3p+q=2+4n'$ where $n$ and $n'$ are integers.
Solving these two equations we get $p=2(n'-n)$  which implies that $p$
is even since $n$ and $n'$ are integers.
From $p+q = 2+4n$, we can say that $p+q$ is even and since $p$ is even
hence $q$ is also even. The assignment $R(X)=0$ then implies $R(Y)=2$ which would allow a $YH_uH_d$ term which would be an intermediate scale $\mu$ term
which we do not allow. Also, for $R(X)=2$ then $R(Y)=0$ and $XH_uH_d$
is allowed and we gain an intermediate scale $\mu$. Hence, we deem no
assignment of $R(X,Y)$ gives viable phenomenology. 

We have run through the complete set of $R(X,Y)$ $R$-charges for
each base model under each discrete $R$-symmetry. Our final tabulation is shown
in Table \ref{tab:RXY}. We find 7 possibilities for base model $B_I$,
10 for $B_{II}$, 21 for $B_{III}$ and 16 for $B_{IV}$, for a total of
54 models.
\begin{table}[h!]
\centering
\resizebox{\textwidth}{!}{\begin{tabular}{|l|l|l|l|l|l|}
\hline
\multicolumn{1}{|c|}{Combination} & $Z_4^R$     & $Z_6^R$                       & $Z_8^R$     & $Z_{12}^R$                              & $Z_{24}^R$                                   \\ \hline
$B_I$                                & none        & (2,2) (5,5)                   & (2,4)       & (2,8) (8,2)                             & (2,20) (14,8)                                \\ \hline
$B_{II}$                               & (1,3) (3,1) & (2,2) (5,5)                   & (3,1) (7,5) & (5,11) (11,5)                           & (11,17) (23,5)                               \\ \hline
$B_{III}$                              & (1,3) (3,1) & (0,2) (2,2) (3,5) (1,5) (5,5) & (1,7) (5,3) & (3,5) (1,11) (7,5) (5,11) (11,5) (9,11) & (5,11) (1,23) (13,11) (9,23) (21,11) (17,23) \\ \hline
$B_{IV}$                               & (3,0) (1,0) & (2,2) (2,5) (5,2) (5,5)       & (7,2) (3,2) & (5,2) (5,8) (11,2) (11,8)               & (11,2) (11,14) (23,2) (23, 14)               \\ \hline
\end{tabular}}
\caption{Allowed $R$-charge assignments for $(X,Y)$ fields for different base
  models and assumed $\mathbb{Z}_n^R$ discrete $R$-symmetries.
  These charges give rise to $\mu\sim m_{weak}\sim \lambda_\mu f_a^2/m_P$.
}
\label{tab:RXY}
\end{table}

\subsection{Operator suppression factors for various base models and $\mathbb{Z}_n^R$ symmetries}
\label{ssec:op}

Next, we wish to tabulate the expected operator suppression factors for
different superpotential combinations listed in Table \ref{tab:QQQ}.
The $R$-charge for trilinear RPV terms under the various $\mathbb{Z}_n^R$
symmetries are listed in row 1 of Table \ref{tab:QQQ} while the bilinear RPV
charges are listed in row 2. Row 3 shows the $R$-charge of the
Weinberg operator $LH_uLH_u$ which turns out to always be $+2$ as expected,
so neutrino masses are always allowed.
In row 4 we show the $R$-charges of the dim-5 proton decay operators
which are always zero. Thus, these dangerous $p$-decay operators will
always receive some suppression when coupled with the $X$ and $Y$ fields.
\begin{table}[!htb]
\renewcommand{\arraystretch}{1.2}
\begin{center}
\begin{tabular}{c|ccccc}
combination & $\mathbb{Z}_{4}^R$ & $\mathbb{Z}_{6}^R$ & $\mathbb{Z}_{8}^R$ & $\mathbb{Z}_{12}^R$ & $\mathbb{Z}_{24}^R$ \\
\hline
$LLE^c$,$LQD^c$, $U^cD^cD^c$  & 3  & 5  & 3 & 11 & 23 \\
$LH_u$ & 1 & 1 & 5 & 1 & 1 \\
$LH_uLH_u$ & 2 & 2 & 2 & 2 & 2 \\
$QQQL$, $U^cU^cD^cE^c$ & 0 & 0 & 0 & 0 & 0 \\
\hline
\end{tabular}
\caption{Derived MSSM field $R$ charge assignments for several
  superpotential operators for various anomaly-free 
discrete $\mathbb{Z}_{n}^R$ symmetries which are consistent with $SU(5)$ or 
$SO(10)$ unification.
}
\label{tab:QQQ}
\end{center}
\end{table}

Next, we check the  operator suppression expected in the $B_I$ base model.
In Table \ref{tab:BIsuppress}, we list in column 1 the base model $B_I$ with
each possibility of $R(X,Y)$ values leading to a weak scale $\mu$ term
and in column 2 we list the associated $\mathbb{Z}_n^R$ symmetry.
In columns 3-5, we list the expected operator suppression by coupling
each operator to the $X$ and $Y$ fields, where $v_X,\ v_Y\sim f_a$.
The first row $B_I$ under $\mathbb{Z}_4^R$ is listed as non-applicable
$(N/A)$ since there are no viable $R(X,Y)$ values leading to $\mu\sim m_{weak}$.
For the second row $B_I(2,2)$ under $\mathbb{Z}_6^R$, since the $R$-charges of
$LH_u$ and $QQQ$ are always odd and $R(X,Y)$ always even, then no combination
of $X$ and $Y$ fields with $LH_u$ or $QQQ$ is allowed and the model has
bRPC and tRPC. However, with $R(QQQQ/m_P)=0$, then $XQQQQ/m_P$ and $YQQQQ/m_P$
are allowed so that when VEVs $v_{X,Y}\sim f_a$ obtain, then we expect
an operator suppression of $f_a/m_P$ which is $\sim 10^{-7}$ for
$f_A\sim 10^{11}$ GeV. The remaining models all have bRPC and tRPC except
$B_I(5,5)$ where bRPV could be allowed at the $\sim 10^{-17}$ GeV level and
tRPV allowed but with a suppression $(f_a/m_P)^3\sim 10^{-21}$.
The latter suppression leads to tRPV couplings of $\sim 10^{-21}$ and so
a lifetime $\tau_{\tchi}\gg \tau_u$, where $\tau_u\sim 4\times 10^{17}$ s
is the age of the universe.
Thus, the strong suppression or non-existence of tRPV terms in
Table \ref{tab:BIsuppress} would bring the (nearly) stable WIMP LSP
into possible conflict with sharp new WIMP direct detection search
limits from LZ\cite{LZ:2024zvo}.
On a more positive note, the dimension-5 proton decay operators are all
suppressed at a level of $10^{-7}$ or well-below.
This is the sort of suppression needed to bring these operators into accord
with experimental limits.
\begin{table}[!htb]
\renewcommand{\arraystretch}{1.2}
\begin{center}
\begin{tabular}{c|c|ccc}
base model & $\mathbb{Z}_{n}^R$ & $LH_u$ & $QQQ$ & $QQQQ/m_P$ \\
\hline
$B_I (X,Y)$ & $\mathbb{Z}_{4}^R$ & N/A & N/A & N/A \\
$B_I (2,2)$ & $\mathbb{Z}_{6}^R$ & bRPC & tRPC & $f_a/m_P$ \\
$B_I (5,5)$ & $\mathbb{Z}_{6}^R$ & $f_a(f_a/m_P)^4$ & $(f_a/m_P)^3$ & $(f_a/m_P)^4$ \\
$B_I (2,4)$ & $\mathbb{Z}_{8}^R$ & bRPC & tRPC & $f_a/m_P$ \\
$B_I (2,8)$ & $\mathbb{Z}_{12}^R$ & bRPC & tRPC & $f_a/m_P$ \\
$B_I (8,2)$ & $\mathbb{Z}_{12}^R$ & bRPC & tRPC & $f_a/m_P$ \\
$B_I (2,20)$ & $\mathbb{Z}_{24}^R$ & bRPC & tRPC & $f_a/m_P$ \\
$B_I (14,8)$ & $\mathbb{Z}_{24}^R$ & bRPC & tRPC & $(f_a/m_P)^4$ \\
\hline
\end{tabular}
\caption{Coefficient suppression of bilinear and trilinear RPV operators
  and dim-5 p-decay operators for base model $B_{I}$ for
  various anomaly-free discrete $\mathbb{Z}_{n}^R$ symmetries
  and for different $(X,Y)$ R-charges which yield $\mu\sim m_{weak}$.
  }
\label{tab:BIsuppress}
\end{center}
\end{table}

In Table \ref{tab:BIIsuppress}, we show the corresponding expected operator
suppression for the ten different versions of base model $B_{II}$.
In this case, the tRPV terms are typically expected to occur with
$\lambda\sim (f_a/m_P)^3\sim 10^{-21}$ level, so the $\tau_{\tchi}> \tau_u$.
However, as noted in Ref. \cite{Baer:2025oid}, several models
$B_{II}(1,3)$, $B_{II}(3,1)$ and $B_{II}(7,5)$ have tRPV at the
$\lambda\sim f_a/m_P\sim 10^{-7}$ level. For this magnitude of tRPV couplings,
then the LSP lifetime is of order $\tau_{\tchi}\sim 10^{-3}-10$ s.
For these sorts of values, then the $\tchi$ still escapes the detector
as MET, and so higgsino pair production should remain quasi-visible at LHC.
But in the early universe, the $\tchi$ can still be produced thermally,
but will all decay away before or during the onset of BBN, leaving
all axion dark matter from SUSY. This seems to be in accord with LZ results,
which find so far no evidence for WIMP dark matter.
One problem with this approach is that $\lambda\sim 10^{-7}$ all by itself is
insufficient to obey the strong tRPV constraints from proton decay. 
Another problem is that the bRPV operators are allowed at the level of
$\mu_i\sim 10^{11}$ GeV, far beyond bounds from $m_{\nu}$ Eq. \ref{eq:mu_i}.
Thus, for this sort of {\it all axion dark matter} scenario to ensue,
then some additional suppression is needed, such as lepton
triality\cite{Ibanez:1991pr}. But such suppressions are almost always needed
in hypotheses of RPV, where usually just one or a few of the
$\lambda, \lambda^\prime, \lambda^{\prime\prime}$ couplings are assumed to be
large. As in the case of base model $B_I$, the dim-5 proton decay operators
are all suppressed in the case of $B_{II}$.
\begin{table}[!htb]
\renewcommand{\arraystretch}{1.2}
\begin{center}
\begin{tabular}{c|c|ccc}
base model & $\mathbb{Z}_{n}^R$ & $LH_u$ & $QQQ$ & $QQQQ/m_P$ \\
\hline
$B_{II} (1,3)$ & $\mathbb{Z}_{4}^R$ & $\sim f_a$  & ${\bf f_a/m_P}$ & $(f_a/m_P)^2$ \\
$B_{II} (3,1)$ & $\mathbb{Z}_{4}^R$ & $\sim f_a$  & ${\bf f_a/m_P}$ & $(f_a/m_P)^2$ \\
$B_{II} (2,2)$ & $\mathbb{Z}_{6}^R$ & bRPC & tRPC & $f_a/m_P$ \\
$B_{II} (5,5)$ & $\mathbb{Z}_{6}^R$ & $f_a(f_a/m_P)^4$ & $(f_a/m_P)^3$ & $(f_a/m_P)^4$\\
$B_{II} (3,1)$ & $\mathbb{Z}_{8}^R$ & $f_a(f_a/m_P)^2$ & $(f_a/m_P)^3$ & $(f_a/m_P)^2$\\
$B_{II} (7,5)$ & $\mathbb{Z}_{8}^R$ & $\sim f_a$ & ${\bf f_a/m_P}$ & $(f_a/m_P)^2$\\
$B_{II} (5,11)$ & $\mathbb{Z}_{12}^R$ & $f_a(f_a/m_P)^4$ & $(f_a/m_P)^3$ & $(f_a/m_P)^4$\\
$B_{II} (11,5)$ & $\mathbb{Z}_{12}^R$ & $f_a(f_a/m_P)^4$ & $(f_a/m_P)^3$ & $(f_a/m_P)^4$\\
$B_{II} (11,17)$ & $\mathbb{Z}_{24}^R$ & $f_a(f_a/m_P)^4$ & $(f_a/m_P)^3$ & $(f_a/m_P)^4$\\
$B_{II} (23,5)$ & $\mathbb{Z}_{24}^R$ & $f_a(f_a/m_P)^4$ & $(f_a/m_P)^3$ & $(f_a/m_P)^4$\\
\hline
\end{tabular}
\caption{Coefficient suppression of bilinear and trilinear RPV operators
  and dim-5 p-decay operators for base model $B_{II}$ for
  various anomaly-free discrete $\mathbb{Z}_{n}^R$ symmetries
  and for different $(X,Y)$ R-charges which yield $\mu\sim m_{weak}$.
The cases which lead to all axion dark matter are labeled in bold.
}
\label{tab:BIIsuppress}
\end{center}
\end{table}

In Table \ref{tab:BIIIsuppress}, we show expected operator suppression
for the 24 versions of base model $B_{III}$. Most cases have both bRPV and
tRPV suppressed either completely or at a high order.
The dimension-5 p-decay operators are also typically highly suppressed.
However, of interest here are base models $B_{III}(3,5)$
(under $\mathbb{Z}_6^R$), $B_{III}(1,7)$ and $B_{III}(3,5)$
(under $\mathbb{Z}_{12}^R$). In these models, the tRPV terms are all suppressed
  at the $f_a/m_P$ level leading to all axion DM, but the bRPV terms
  appear at the $f_a(f_a/m_P)^2\sim 1$ MeV level.
  The latter results are apparently in accord with the $m_\nu$ constraints on
  bRPV terms. In these cases, the dim-5 $p$-decay operators are suppressed
  at the $(f_a/m_P)^2\sim 10^{-14}$ level. For these cases, baryon triality or
  hexality could be invoked to bring the tRPV terms into accord with
  $p$-decay constraints.
\begin{table}[!htb]
\renewcommand{\arraystretch}{1.2}
\begin{center}
\begin{tabular}{c|c|ccc}
base model & $\mathbb{Z}_{n}^R$ & $LH_u$ & $QQQ$ & $QQQQ/m_P$ \\
\hline
$B_{III} (1,3)$ & $\mathbb{Z}_{4}^R$ & $\sim f_a$  & ${\bf f_a/m_P}$ & $(f_a/m_P)^2$ \\
$B_{III} (3,1)$ & $\mathbb{Z}_{4}^R$ & $\sim f_a$  & ${\bf f_a/m_P}$ & $(f_a/m_P)^2$ \\
$B_{III} (0,2)$ & $\mathbb{Z}_{6}^R$ & bRPC & tRPC & $f_a/m_P$ \\
$B_{III} (2,2)$ & $\mathbb{Z}_{6}^R$ & bRPC & tRPC & $f_a/m_P$ \\
$B_{III} (3,5)$ & $\mathbb{Z}_{6}^R$ & $f_a(f_a/m_P)^2$ & ${\bf f_a/m_P}$ & $(f_a/m_P)^2$ \\
$B_{III} (1,5)$ & $\mathbb{Z}_{6}^R$ & $\sim f_a$ & $(f_a/m_P)^3$ & $(f_a/m_P)^2$ \\
$B_{III} (5,5)$ & $\mathbb{Z}_{6}^R$ & $f_a(f_a/m_P)^4$ & $(f_a/m_P)^3$ & $(f_a/m_P)^4$ \\
$B_{III} (1,7)$ & $\mathbb{Z}_{8}^R$ & $f_a(f_a/m_P)^2$ & ${\bf f_a/m_P}$ & $(f_a/m_P)^2$ \\
$B_{III} (5,3)$ & $\mathbb{Z}_{8}^R$ & $\sim f_a$ & $(f_a/m_P)^3$ & $(f_a/m_P)^2$ \\
$B_{III} (3,5)$ & $\mathbb{Z}_{12}^R$ & $f_a(f_a/m_P)^2$ & ${\bf f_a/m_P}$ & $(f_a/m_P)^4$ \\
$B_{III} (1,11)$ & $\mathbb{Z}_{12}^R$ & $\sim f_a$ & $(f_a/m_P)^3$ & $(f_a/m_P)^2$ \\
$B_{III} (7,5)$ & $\mathbb{Z}_{12}^R$ & $f_a(f_a/m_P)^4$ & $(f_a/m_P)^3$ & $(f_a/m_P)^2$ \\
$B_{III} (5,11)$ & $\mathbb{Z}_{12}^R$ & $f_a(f_a/m_P)^4$ & $(f_a/m_P)^3$ & $(f_a/m_P)^4$ \\
$B_{III} (11,5)$ & $\mathbb{Z}_{12}^R$ & $f_a(f_a/m_P)^4$ & $(f_a/m_P)^3$ & $(f_a/m_P)^4$ \\
$B_{III} (9,11)$ & $\mathbb{Z}_{12}^R$ & $f_a(f_a/m_P)^4$ & $(f_a/m_P)^3$ & $(f_a/m_P)^4$ \\
$B_{III} (5,11)$ & $\mathbb{Z}_{24}^R$ & $f_a(f_a/m_P)^4$ & $(f_a/m_P)^3$ & $(f_a/m_P)^4$ \\
$B_{III} (1,23)$ & $\mathbb{Z}_{24}^R$ & $\sim f_a$ & $(f_a/m_P)^3$ & $(f_a/m_P)^2$ \\
$B_{III} (13,11)$ & $\mathbb{Z}_{24}^R$ & $f_a(f_a/m_P)^{10}$ & $(f_a/m_P)^9$ & $(f_a/m_P)^2$ \\
$B_{III} (9,23)$ & $\mathbb{Z}_{24}^R$ & $f_a(f_a/m_P)^4$ & $(f_a/m_P)^3$ & $(f_a/m_P)^4$ \\
$B_{III} (21,11)$ & $\mathbb{Z}_{24}^R$ & $f_a(f_a/m_P)^6$ & $(f_a/m_P)^5$ & $(f_a/m_P)^4$ \\
$B_{III} (17,23)$ & $\mathbb{Z}_{24}^R$ & $f_a(f_a/m_P)^4$ & $(f_a/m_P)^3$ & $(f_a/m_P)^4$ \\
\hline
\end{tabular}
\caption{Coefficient suppression of bilinear and trilinear RPV operators
  and dim-5 p-decay operators for base model $B_{III}$ for
  various anomaly-free discrete $\mathbb{Z}_{n}^R$ symmetries
  and for different $(X,Y)$ R-charges which yield $\mu\sim m_{weak}$.
  }
\label{tab:BIIIsuppress}
\end{center}
\end{table}

In Table \ref{tab:BIVsuppress}, we list expected operator suppression for
the 16 versions of base model $B_{IV}$. The bulk of the $B_{IV}$ models
have tRPV either highly or completely suppressed. Also, dim-5 $p$-decay
is also well-suppressed. However, in this case, the models
$B_{IV}(3,0)$ and $B_{IV}(7,2)$ have tRPV suppression at the $f_a/m_P$ level
and so would lead to all axion dark matter. In addition, like the $B_{III}$
models, these also have sufficient suppression of bRPV terms.
Some additional suppression such as baryon or lepton triality or proton
hexality\cite{Dreiner:2005rd} would be needed as usual to bring the
$\lambda^\prime\lambda^{\prime\prime}$ product into accord with the tight $p$-decay 
constraints.
\begin{table}[!htb]
\renewcommand{\arraystretch}{1.2}
\begin{center}
\begin{tabular}{c|c|ccc}
base model & $\mathbb{Z}_{n}^R$ & $LH_u$ & $QQQ$ & $QQQQ/m_P$ \\
\hline
$B_{IV} (1,0)$ & $\mathbb{Z}_{4}^R$ & $\sim f_a$  & $(f_a/m_P)^3$ & $(f_a/m_P)^2$ \\
$B_{IV} (3,0)$ & $\mathbb{Z}_{4}^R$ & $f_a(f_a/m_P)^2$  & ${\bf f_a/m_P}$ & $(f_a/m_P)^2$ \\
$B_{IV} (2,2)$ & $\mathbb{Z}_{6}^R$ & bRPC  & tRPC & $f_a/m_P$ \\
$B_{IV} (2,5)$ & $\mathbb{Z}_{6}^R$ & $f_a(f_a/m_P)$  & $(f_a/m_P)^3$ & $f_a/m_P$ \\
$B_{IV} (5,2)$ & $\mathbb{Z}_{6}^R$ & $f_a(f_a/m_P)$  & $(f_a/m_P)^3$ & $f_a/m_P$ \\
$B_{IV} (5,5)$ & $\mathbb{Z}_{6}^R$ & $f_a(f_a/m_P)^4$  & $(f_a/m_P)^3$ & $(f_a/m_P)^4$ \\
$B_{IV} (3,2)$ & $\mathbb{Z}_{8}^R$ & $f_a(f_a/m_P)$  & $(f_a/m_P)^3$ & $f_a/m_P$ \\
$B_{IV} (7,2)$ & $\mathbb{Z}_{8}^R$ & $f_a(f_a/m_P)^2$  & ${\bf f_a/m_P}$ & $f_a/m_P$ \\
$B_{IV} (5,2)$ & $\mathbb{Z}_{12}^R$ & $f_a(f_a/m_P)^4$  & $(f_a/m_P)^3$ & $f_a/m_P$ \\
$B_{IV} (5,8)$ & $\mathbb{Z}_{12}^R$ & $f_a(f_a/m_P)$  & $(f_a/m_P)^3$ & $(f_a/m_P)^4$ \\
$B_{IV} (11,2)$ & $\mathbb{Z}_{12}^R$ & $f_a(f_a/m_P)$  & $(f_a/m_P)^3$ & $f_a/m_P$ \\
$B_{IV} (11,8)$ & $\mathbb{Z}_{12}^R$ & $f_a(f_a/m_P)^4$  & $(f_a/m_P)^3$ & $(f_a/m_P)^4$ \\
$B_{IV} (11,2)$ & $\mathbb{Z}_{24}^R$ & $f_a(f_a/m_P)^7$  & $(f_a/m_P)^9$ & $f_a/m_P$ \\
$B_{IV} (11,14)$ & $\mathbb{Z}_{24}^R$ & $f_a(f_a/m_P)$  & $(f_a/m_P)^6$ & $(f_a/m_P)^4$ \\
$B_{IV} (23,2)$ & $\mathbb{Z}_{24}^R$ & $f_a(f_a/m_P)$  & $(f_a/m_P)^3$ & $f_a/m_P$ \\
$B_{IV} (23,14)$ & $\mathbb{Z}_{24}^R$ & $f_a(f_a/m_P)^4$  & $(f_a/m_P)^3$ & $(f_a/m_P)^4$ \\
\hline
\end{tabular}
\caption{Coefficient suppression of bilinear and trilinear RPV operators
  and dim-5 p-decay operators for base model $B_{IV}$ for
  various anomaly-free discrete $\mathbb{Z}_{n}^R$ symmetries
  and for different $(X,Y)$ R-charges which yield $\mu\sim m_{weak}$.
  }
\label{tab:BIVsuppress}
\end{center}
\end{table}

\section{Tree-level LSP decay modes and decay rates}
\label{sec:LSPdecay}

In this Section, we examine in more detail the LSP decay modes engendered
by non-zero tRPV couplings in Eq. \ref{eq:Wrpv}. In Ref. \cite{Baer:2025oid},
a simple approximate expression was used for the LSP decay rates which was
suspected to be an order-of-magnitude estimate: for a photino-like $\tchi_1^0$
state decaying via a $\lambda_{ijk}^\prime$ coupling, then
\be
\Gamma (\tchi_1^0 )= \frac{3\alpha\lambda_{ijk}^{\prime 2}}{128\pi^2}
\frac{m_{\tchi_1^0}^5}{m_{soft}^4}
\label{eq:Gamma}
\ee
where $m_{soft}$ is indicative of the virtual sfermion being exchanged
in the decay process. In reality, the lightest neutralino is a
gaugino-higgsino admixture and the decay depends on the neutralino gaugino and
higgsino mixing angles and the relevant Yukawa coupling which enters the
higgsino mixing contribution to the coupling.\footnote{See {\it e.g.}
  pages 167-170 of Ref. \cite{Baer:2006rs} for neutralino couplings
  including all mixing angle, gauge and Yukawa coupling factors.}

\begin{figure}[htb!]
  \centering
    % (a) mediated by \tilde t
    \begin{subfigure}{0.49\textwidth}
      \centering
      \includegraphics[width=0.8\textwidth]{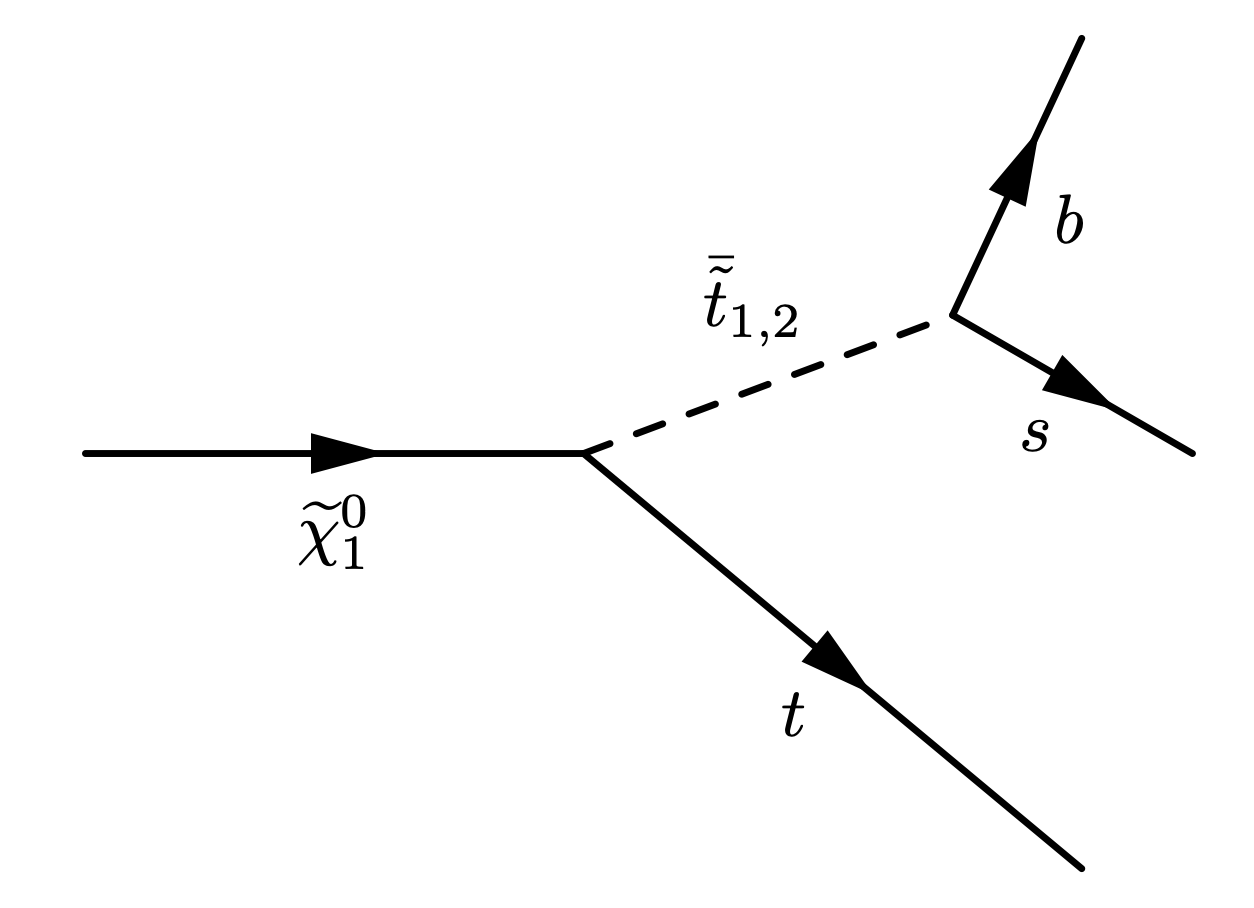}
      \caption{$\tilde t_{1,2}$‐mediated}\label{fig:decay_t}
    \end{subfigure}
    \hfill
    % (b) mediated by \tilde s
    \begin{subfigure}{0.49\textwidth}
      \centering
      \includegraphics[width=0.8\textwidth]{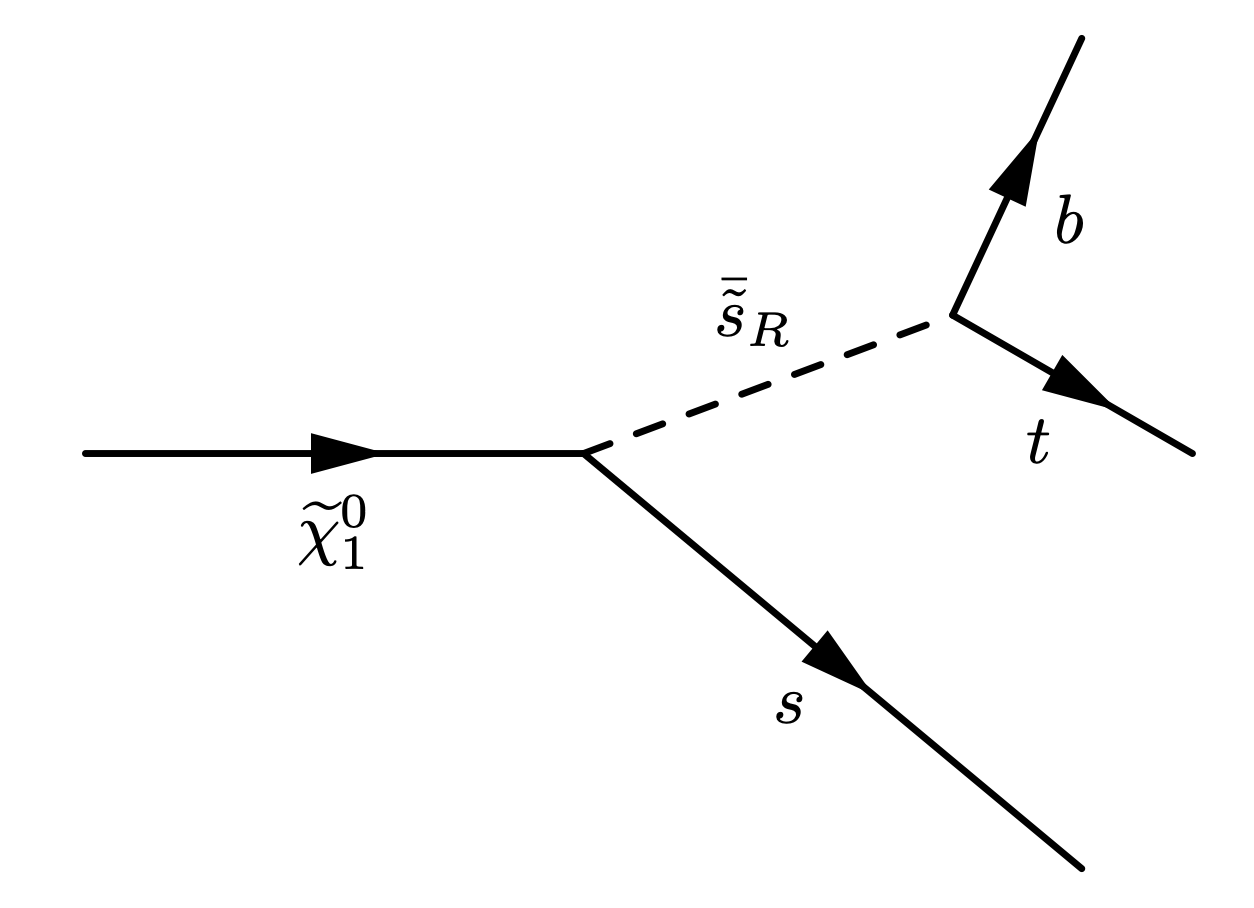}
      \caption{$\tilde s_R$‐mediated}\label{fig:decay_s}
    \end{subfigure}
    \hfill \\
    % (c) mediated by \tilde b
    \begin{subfigure}{0.49\textwidth}
      \centering
      \includegraphics[width=0.8\textwidth]{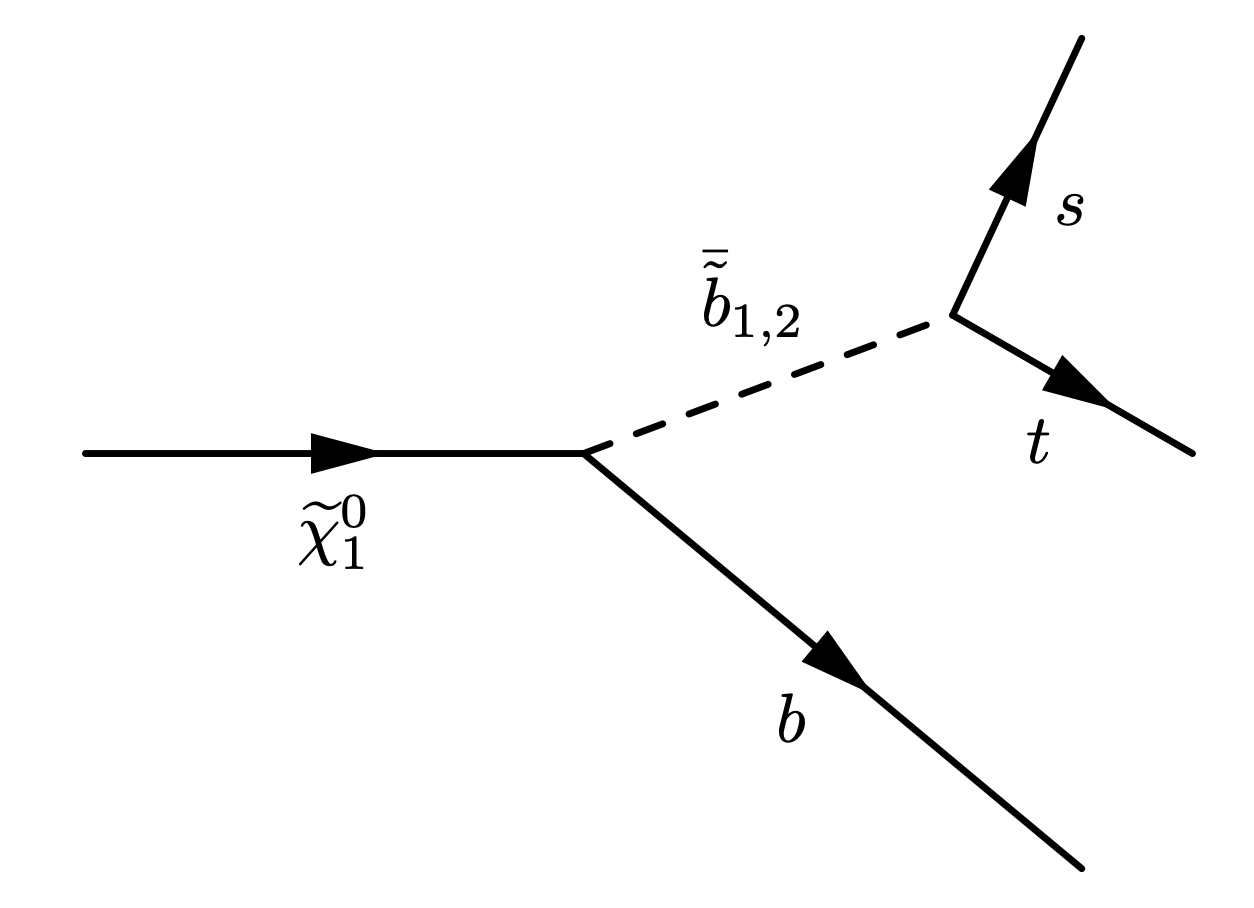}
      \caption{$\tilde b_{1,2}$‐mediated}\label{fig:decay_b}
    \end{subfigure}
  \caption{Tree‐level diagrams for the RPV decay
    $\tilde\chi^0_1\!\to t\,s\,b$ via off‐shell squark exchange:
    (a) $\tilde t_{1,2}$‐mediated, (b) $\tilde s_R$‐mediated, and (c) $\tilde b_{1,2}$‐mediated.}
  \label{fig:diagrams}
\end{figure}

Also, the decays can take place via several different 
intermediate-state sfermions which may have very different mass scales:
for instance, in landscape SUSY models, the first/second generation sfermions
are pulled to the 10-40 TeV level whilst third generation sfermions typically
occur at the 1-10 TeV level\cite{Baer:2017uvn}.
As an example, the five decay diagrams for the case where $\tchi_1^0\to tbs$
via the $\lambda_{323}^{\prime\prime}$ coupling are shown in
Fig. \ref{fig:diagrams}, generated by \textsc{MadGraph5}.
From the Figure, we see that the diagrams depend
on each of the $\tchi_1^0 b\tb_1$, $\tchi_1^0 b\tb_2$,
$\tchi_1^0 t\tst_1$, $\tchi_1^0 t\tst_2^0$ and $\tchi_1^0 s\ts_R$
vertices, each of which is different.
The diagrams also depend on the sfermion mixing angles through the
tRPV couplings. It can also be important to include phase space effects
when the final states include top- or bottom-quarks.
%
%\begin{figure}[htb!]
%\centering
%    {\includegraphics[height=0.8\textheight]{diagrams_1_n1_tsb.pdf}
%      \caption{Tree-level Feynman diagrams for lightest neutralino decay
%$\tchi_1^0\to tbs$ via the $\lambda^{\prime\prime}_{ijk}$ couplings.}
%     \label{fig:diagrams}}
%\end{figure}
%

Since our operator suppression analysis implies that {\it all}
tRPV couplings are comparably suppressed, it can also be important to include
all possible decay final states.
In Table \ref{tab:lambda}, we list the complete set of $\tchi_1^0$
final states allowed under the $\lambda_{ijk}$ tRPV couplings.
There are four distinct final states for each of the nine $\lambda_{ijk}$
couplings, corresponding to 36 different decay modes. Since the final state
particles are all quite light compared to $m_{\tchi_1^0}$, then kinematic
effects are not so important here.
\begin{table}[!htb]
\renewcommand{\arraystretch}{1.2}
\begin{center}
\begin{tabular}{c|c}
coupling & $\tchi_1^0$ decay modes \\
\hline
$\lambda_{121}$ & $e\bar{\nu}_e\bar{\mu}+\bar{e}\nu_e\mu+
\bar{\nu}_\mu e\bar{e}+\nu_\mu e\bar{e}$ \\
$\lambda_{122}$ & $\mu\bar{\nu}_e\bar{\mu}+\bar{\mu}\nu_e\mu +
\bar{\nu}_\mu\bar{e} e +\nu_\mu e\bar{\mu}$ \\
$\lambda_{123}$ & $\tau\bar{\nu}_e\bar{\mu}+\bar{\tau}\nu_e\mu +
\bar{\nu}_\mu\bar{e}\tau +\nu_\mu e\bar{\tau}$ \\
$\lambda_{131}$ & $e\bar{\nu}_e\bar{\tau}+\bar{e}\nu_e\tau +
\bar{\nu}_\tau \bar{e} e+ \nu_\tau e\bar{e}$ \\
$\lambda_{132}$ & $\mu\bar{\nu}_e\bar{\tau} +\bar{\mu}\nu_e\tau +
\bar{\nu}_\tau\bar{e}\mu +\nu_\tau e\bar{\mu}$ \\
$\lambda_{133}$ & $\tau\bar{\nu}_e\bar{\tau} +\bar{\tau}\nu_e\tau +
\bar{\nu}_\tau\bar{e}\tau +\nu_\tau e\bar{\tau}$ \\
$\lambda_{231}$ & $\bar{\nu}_\mu\bar{\tau}e +\nu_\mu\tau\bar{e} +
\bar{\nu}_\tau\bar{\mu} e+ \nu_\tau \mu\bar{e}$ \\
$\lambda_{232}$ & $\bar{\nu}_\mu\bar{\tau}\mu +\nu_\mu\tau\bar{\mu} +
\bar{\nu}_\tau\bar{\mu}\mu +\nu_\tau \mu\bar{\mu}$ \\
$\lambda_{233}$ & $\bar{\nu}_\mu\bar{\tau}\tau +\nu_\mu\tau\bar{\tau} +
\bar{\nu}_\tau\bar{\mu}\tau +\nu_\tau\mu\bar{\tau}$ \\
\hline
\end{tabular}
\caption{RPV couplings $\lambda_{ijk}$ along with 36 lightest neutralino
  $\tchi_1^0$ decay modes engendered by these couplings.
}
\label{tab:lambda}
\end{center}
\end{table}

In Table \ref{tab:lambda_p}, we list all four final states allowed for
each of the 27 $\lambda_{ijk}^\prime$ couplings.
Thus, from the Table, we see 108 distinct final states from
$\tchi_1^0$ via the $\lambda_{ijk}^\prime$ couplings.
These couplings lead to mixed hadronic-leptonic final states which
can produce very different collider signatures in the case where
$\tchi_1^0$ decays within the detector geometry.
\begin{table}[!htb]
\renewcommand{\arraystretch}{1.2}
\begin{center}
\begin{tabular}{c|c}
coupling & $\tchi_1^0$ decay modes \\
\hline
$\lambda^{\prime}_{111}$ & $\bar{\nu}_e \bar{d}d+\bar{e}\bar{u}d+\nu_e d\bar{d}+e u\bar{d}$ \\
$\lambda^{\prime}_{112}$ & $\bar{\nu}_e \bar{d}s+\bar{e}\bar{u}\bar{s}+\nu_e d\bar{s}+e u\bar{s}$ \\ 
$\lambda^{\prime}_{113}$ & $\bar{\nu}_e \bar{d}b+\bar{e}\bar{u}b+\nu_e d\bar{b}+e u\bar{b}$ \\ 
$\lambda^{\prime}_{121}$ & $\bar{\nu}_e \bar{s}d+\bar{e}\bar{c}d+\nu_e s\bar{d}+e c\bar{d}$ \\
$\lambda^{\prime}_{122}$ & $\bar{\nu}_e s\bar{s}+\bar{e}\bar{c}s+\nu_e s\bar{s}+e c\bar{s}$ \\
$\lambda^{\prime}_{123}$ & $\bar{\nu}_e \bar{s}b+\bar{e}\bar{c}b+\nu_e s\bar{b}+e c\bar{b}$ \\
$\lambda^{\prime}_{131}$ & $\bar{\nu}_e \bar{b}d+\bar{e}\bar{t}d+\nu_e b\bar{d}+e t\bar{d}$ \\
$\lambda^{\prime}_{132}$ & $\bar{\nu}_e \bar{b}s+\bar{e}\bar{t}\bar{s}+\nu_e b\bar{s}+e t\bar{s}$ \\
$\lambda^{\prime}_{133}$ & $\bar{\nu}_e b\bar{b}+\bar{e}\bar{t}b+\nu_e b\bar{b}+e t\bar{b}$ \\

$\lambda^{\prime}_{211}$ & $\bar{\nu}_\mu \bar{d}d+\bar{\mu}\bar{u}d+\nu_\mu \bar{d}d+\mu u\bar{d}$ \\
$\lambda^{\prime}_{212}$ & $\bar{\nu}_\mu \bar{d}s+\bar{\mu}\bar{u}s+\nu_\mu d\bar{s}+\mu u\bar{s}$ \\
$\lambda^{\prime}_{213}$ & $\bar{\nu}_\mu \bar{d}b+\bar{\mu}\bar{u}b+\nu_\mu d\bar{b}+\mu u\bar{b}$ \\
$\lambda^{\prime}_{221}$ & $\bar{\nu}_\mu \bar{s}d+\bar{\mu}\bar{c}d+\nu_\mu s\bar{d}+\mu c\bar{d}$ \\
$\lambda^{\prime}_{222}$ & $\bar{\nu}_\mu \bar{s}s+\bar{\mu}\bar{c}s+\nu_\mu s\bar{s}+\mu c\bar{s}$ \\
$\lambda^{\prime}_{223}$ & $\bar{\nu}_\mu \bar{s}d+\bar{\mu}\bar{c}b+\nu_\mu s\bar{b}+\mu c\bar{b}$ \\
$\lambda^{\prime}_{231}$ & $\bar{\nu}_\mu \bar{b}d+\bar{\mu}\bar{t}d+\nu_\mu b\bar{d}+\mu t\bar{d}$ \\
$\lambda^{\prime}_{232}$ & $\bar{\nu}_\mu \bar{b}s+\bar{\mu}\bar{t}s+\nu_\mu b\bar{s}+\mu t\bar{s}$ \\
$\lambda^{\prime}_{233}$ & $\bar{\nu}_\mu \bar{b}b+\bar{\mu}\bar{t}b+\nu_\mu b\bar{b}+\mu t\bar{b}$ \\

$\lambda^{\prime}_{311}$ & $\bar{\nu}_\tau \bar{d}d+\bar{\tau}\bar{u}d+\nu_\tau d\bar{d}+\tau u\bar{d}$ \\
$\lambda^{\prime}_{312}$ & $\bar{\nu}_\tau \bar{d}s+\bar{\tau}\bar{u}s+\nu_\tau d\bar{s}+\tau u\bar{s}$ \\
$\lambda^{\prime}_{313}$ & $\bar{\nu}_\tau \bar{d}b+\bar{\tau}\bar{u}b+\nu_\tau d\bar{b}+\tau u\bar{b}$ \\
$\lambda^{\prime}_{321}$ & $\bar{\nu}_\tau \bar{s}d+\bar{\tau}\bar{c}d+\nu_\tau s\bar{d}+\tau c\bar{d}$ \\
$\lambda^{\prime}_{322}$ & $\bar{\nu}_\tau \bar{s}s+\bar{\tau}\bar{c}s+\nu_\tau s\bar{s}+\tau c\bar{s}$ \\
$\lambda^{\prime}_{323}$ & $\bar{\nu}_\tau \bar{s}b+\bar{\tau}\bar{c}b+\nu_\tau s\bar{b}+\tau c\bar{b}$ \\
$\lambda^{\prime}_{331}$ & $\bar{\nu}_\tau \bar{b}d+\bar{\tau}\bar{t}d+\nu_\tau b\bar{d}+\tau t\bar{d}$ \\
$\lambda^{\prime}_{332}$ & $\bar{\nu}_\tau \bar{b}s+\bar{\tau}\bar{t}s+\nu_\tau b\bar{s}+\tau t\bar{s}$ \\
$\lambda^{\prime}_{333}$ & $\bar{\nu}_\tau \bar{b}b+\bar{\tau}\bar{t}b+\nu_\tau b\bar{b}+\tau t\bar{b}$ \\

\hline
\end{tabular}
\caption{RPV couplings $\lambda^{\prime}_{ijk}$ along with 108 lightest neutralino
  $\tchi_1^0$ decay modes engendered by these couplings.
}
\label{tab:lambda_p}
\end{center}
\end{table}

In Table \ref{tab:lambda_pp}, we show the $\tchi_1^0$ decay modes
produced by the 9 $\lambda_{ijk}^{\prime\prime}$.
Here, there are only two distinct final states produced from each
coupling, for a total of 18 different purely hadronic final states.
These final states are not expected to produce large MET since no
neutrinos are involved in the decay processes.
Both $\lambda_{ijk}^\prime$ and $\lambda_{ijk}^{\prime\prime}$ can produce
top-quarks in the final state where phase space effects may be important.
\begin{table}[!htb]
\renewcommand{\arraystretch}{1.2}
\begin{center}
\begin{tabular}{c|c}
coupling & $\tchi_1^0$ decay modes \\
\hline
$\lambda^{\prime\prime}_{112}$ & $uds +\bar{u}\bar{d}\bar{s}$ \\
$\lambda^{\prime\prime}_{113}$ & $udb +\bar{u}\bar{d}\bar{b}$ \\
$\lambda^{\prime\prime}_{123}$ & $usb +\bar{u}\bar{s}\bar{b}$ \\
$\lambda^{\prime\prime}_{212}$ & $cds +\bar{c}\bar{d}\bar{s}$ \\
$\lambda^{\prime\prime}_{213}$ & $cdb +\bar{c}\bar{d}\bar{b}$ \\
$\lambda^{\prime\prime}_{223}$ & $csb +\bar{c}\bar{s}\bar{b}$ \\
$\lambda^{\prime\prime}_{312}$ & $tds +\bar{t}\bar{d}\bar{s}$ \\
$\lambda^{\prime\prime}_{313}$ & $tdb +\bar{t}\bar{d}\bar{b}$ \\
$\lambda^{\prime\prime}_{323}$ & $tsb +\bar{t}\bar{s}\bar{b}$ \\
\hline
\end{tabular}
\caption{RPV couplings $\lambda^{\prime\prime}_{ijk}$ along with 18
  lightest neutralino
  $\tchi_1^0$ decay modes engendered by these couplings.
}
\label{tab:lambda_pp}
\end{center}
\end{table}

The sparticle spectrum and mixing matrices are generated with
\texttt{ISAJET}~\cite{Paige:2003mg}, yielding a predominantly higgsino
lightest neutralino $\tz_1$ in our natural SUSY benchmark point.
The results are exported to a SUSY Les Houches Accord (SLHA)
file\cite{Skands:2003cj}, which is then fed into
\textsc{MadGraph5}~\cite{Alwall:2011uj,Alwall:2014hca}.
We simulate R-parity–violating decays of $\tz_1$ in \textsc{MadGraph5}
using the RPV–MSSM UFO model implementation of Ref.~\cite{RPVUFO}.
For each scan point we pick one independent entry $\lambda''$ of the
SLHA \texttt{RVLAMUDD} block, and also set $\lambda''_{ikj}=-\lambda''_{ijk}$
to enforce antisymmetry with respect to the last two indices.
\textsc{MadGraph5} then automatically calculates all kinematically allowed tree-level
decay amplitudes of $\tz_1$ and integrates over the final state
phase space to obtain numerically the resulting widths.

For our sample benchmark model, we adopt the three-extra-parameter
non-universal Higgs model\cite{Ellis:2002wv} (NUHM3) with parameter space given by
\be
m_0(1,2),\ m_0(3),\ m_{1/2},\ A_0,\ \tan\beta,\ \mu,\ m_A\ \ \ (NUHM3)
\ee
and where we take $m_0(1,2)=30$ TeV, $m_0(3)=6.2$ TeV, $m_{1/2}=2.14$ TeV,
$A_0=-6.2$ TeV, $\tan\beta =10$, $\mu =200$ GeV and $m_A=2$ TeV.
The model has all sparticles above LHC search limits with
$m_{\tg}=5$ TeV and $m_{\tst_1}=1.2$ TeV and with $m_h=124.9$ GeV. 
The model is EW natural with $\Delta_{EW}\sim 30$.

\begin{figure}[htb!]
\centering
    {\includegraphics[height=0.6\textheight]{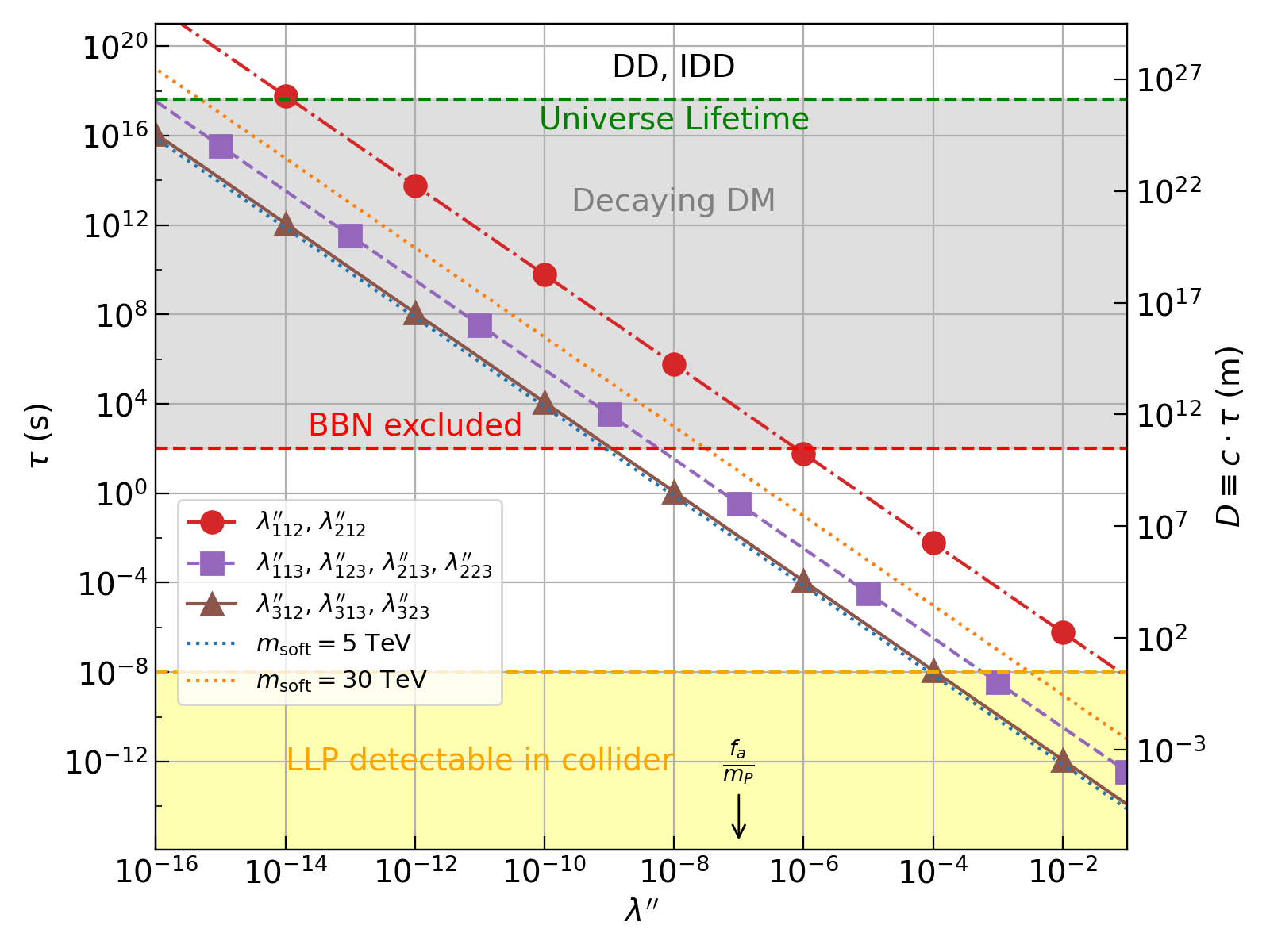}
      \caption{Lightest neutralino lifetime vs. $\lambda_{ijk}^{\prime\prime}$ for
        the natural SUSY benchmark model listed in the text, with
        $m_{\tchi_1^0}\sim 100$ GeV for various $ijk$ values.
        The region labeled DD and IDD is where direct detection and
        indirect detection of WIMPs is relevant.
        }
     \label{fig:tau_z1}}
\end{figure}

Our first results are shown in Fig.~\ref{fig:tau_z1}, where we show the
lifetime $\tau_{\tchi_1^0}$ vs. $\lambda_{ijk}^{\prime\prime}$ for several
different choices of $ijk=112$ and $212$ (red dots), $113$, $123$, $213$ and
$223$ (purple squares) and $312$, $313$ and $323$ (brown triangles).
On the right vertical axis, we show the approximate decay length
$D=c\tau$ in meters. For comparison, we also show the order-of-magnitude
estimate Eq. \ref{eq:Gamma} for $m_{soft}=5$ and 30 TeV as the
blue and yellow dotted lines, respectively.
The yellow-shaded region is where we expect $\tchi_1^0$ to decay within
LHC detector volumes.
The gray-shaded region has decaying $\tchi_1^0$ states but where the decays
violate BBN constraints on late-decaying neutral particles in the early
universe, as computed by Jedamzik\cite{Jedamzik:2006xz}.
The green-dashed line denotes the age of the universe, and $\tchi_1^0$
decay widths above this line can be regarded as stable and susceptible from
WIMP direct detection (DD) and indirect detection (IDD) search limits.

From the plot, we can compare the approximate dotted curves against the exact
tree-level lifetime calculations. We find a spread of results, so that the
exact and approximate results are typically within an order of magnitude.
The exact results for different $\lambda_{ijk}^{\prime\prime}$ are also spread
over four orders of magnitude. The third generation
$\lambda_{312}^{\prime\prime}$, $\lambda_{313}^{\prime\prime}$ and
$\lambda_{323}^{\prime\prime}$ give larger decay widths (lower lifetimes)
due to the large higgsino-component of $\tchi_1^0$, where also the large
top Yukawa coupling enters the decay width.
The $\tchi_1^0$ decay width to first/second generation quarks is thus
suppressed compared to decays to third generation particles.

\begin{figure}[htb!]
\centering
    {\includegraphics[height=0.6\textheight]{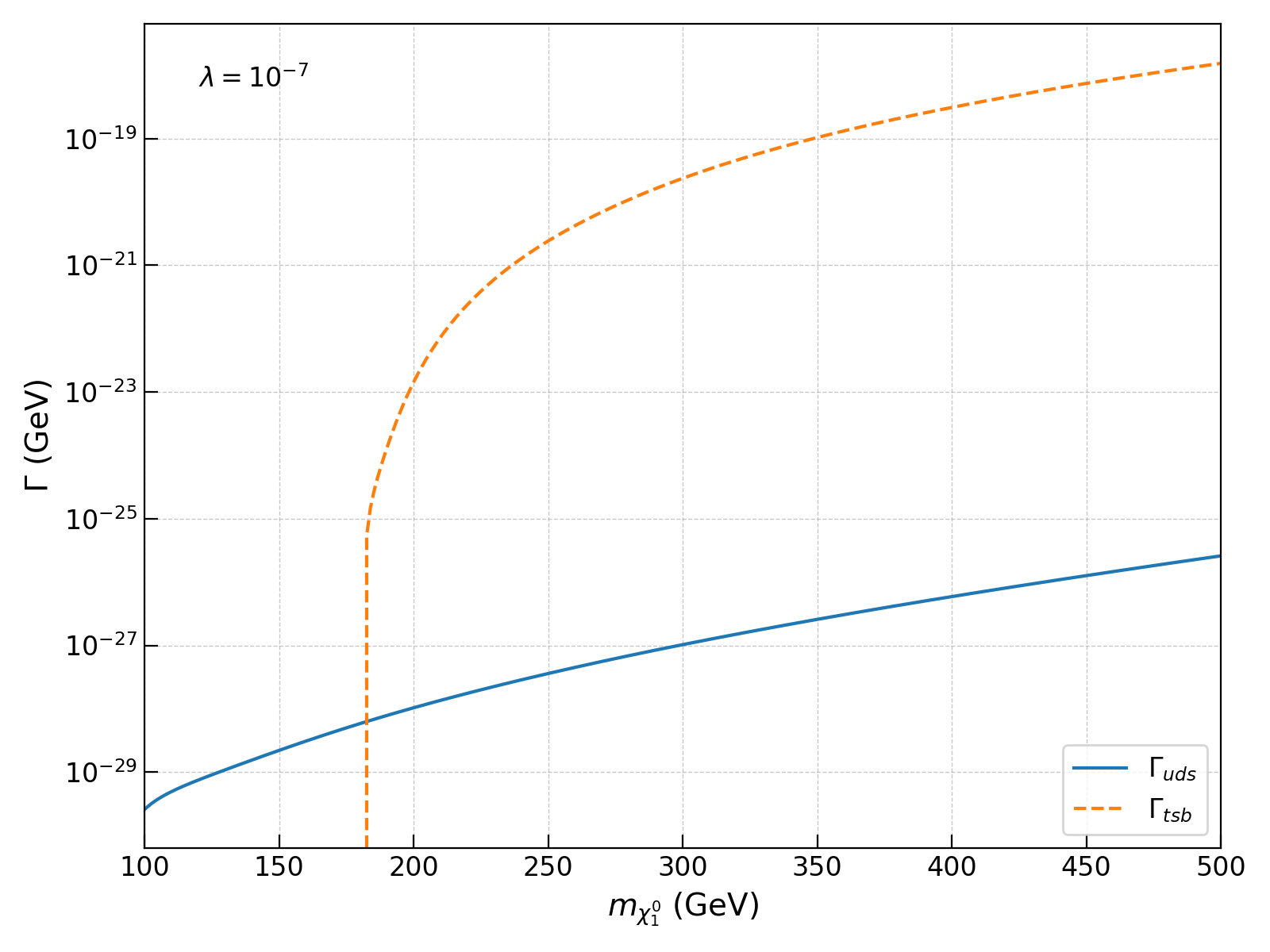}
      \caption{Lightest neutralino lifetime vs. $m_{\tchi_1^0}$ for
        $\lambda_{112}^{\prime\prime}$ and $\lambda_{323}^{\prime\prime} =10^{-7}$ for
        the natural SUSY benchmark model listed in the text.
        }
     \label{fig:tau_mz1}}
\end{figure}

In Fig. \ref{fig:tau_mz1}, we show the $\tchi_1^0$ decay width
$\Gamma (\tchi_1^0\to uds )$ vs. $m_{\tchi_1^0}$ and compare this
to $\Gamma (\tchi_1^0\to tbs )$. From the previous discussion, we
might expect decays to third generation quarks to dominate, and indeed this
is the case for large $m_{\tchi_1^0}$ when phase space effects are
unimportant. But for lower values of $m_{\tchi_1^0}<m_t+m_b+m_s$, then
the decay to third generation quarks is kinematically suppressed or forbidden,
and the decay to first/second generation quarks can be dominant.

\section{Limits on RPV couplings from LHC in natural SUSY}
\label{sec:lhc}

In the textbook construction of the MSSM, a weak-scale SUSY conserving
$\mu$ term is incorporated whilst $R$-parity conservation is invoked to
forbid the RPV terms in Eq. \ref{eq:Wrpv}. Both these steps are ad-hoc,
and should arise from some deeper physics reason.
The aforementioned anomaly-free discrete $R$ symmetries $\mathbb{Z}_n^R$
(consistent with $SU(5)$ or $SO(10)$ GUT conditions)
can arise from string compactification to 4-dimensions and have the
advantage of forbidding the $\mu$ term, the RPV terms in Eq. \ref{eq:Wrpv}
and the dangerous dimension-5 $p$-decay operators.
The $\mu$ parameter can be regenerated at the weak scale by invoking
the Kim-Nilles solution which requires the addition of two or more
singlet fields ($X$ and $Y$) coupled to $H_uH_d$.
The underlying theory then displays an accidental global $U(1)_{PQ}$
which can be used to solve the strong CP problem where the SUSY DFSZ axion
is composed of the mixed $X$ and $Y$ fields.
This makes the MSSM much more plausible. An additional step to plausibility is
to work within EW {\it natural} SUSY, wherein the weak scale is
$m_{W,Z,h}\sim 100$ GeV because all contributions to the weak scale are
comparable to or less than the weak scale (practical naturalness with
naturalness measure $\Delta_{EW}\alt 30$). In natural SUSY models,
the several {\it higgsinos} compose the lightest SUSY particles since $\mu$
is SUSY conserving and feeds mass to $W,Z,h$ and higgsinos.
Thus, a consequence of our effort to be plausible is that we expect light
higgsino-like EWinos with $m_{\tchi_{1,2}^0}$ and $m_{\tchi_1^\pm}\sim 100-350$ GeV. 

\begin{figure}[htb!]
\centering
    {\includegraphics[height=0.6\textheight]{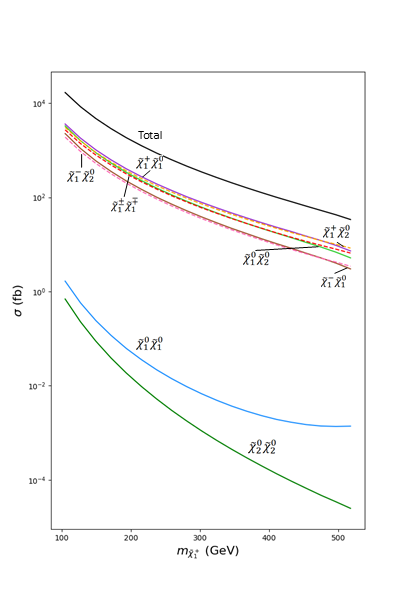}}
    \caption{Cross section for various higgsino pair production reactions
      at LHC14 vs. $m_{\tchi_1^\pm}\simeq \mu$ for a natural SUSY benchmark point.
     \label{fig:sig}}
\end{figure}

In Fig. \ref{fig:sig}, we show the various light higgsino pair production
cross sections at LHC14 ($\sqrt{s}=14$ TeV) versus $m_{\tchi_1^\pm}\sim \mu$.
We use Prospino\cite{Beenakker:1996ed} to generate the NLO pair production
cross sections and we use a natural SUSY benchmark point from the
NUHM2\cite{Ellis:2002wv,Ellis:2002iu,Baer:2005bu} model
(as generated with Isajet\cite{Paige:2003mg}) with parameters $m_0=5$ TeV,
$m_{1/2}=1.2$ TeV, $A_0=-8$ TeV, $\tan\beta =10$ with $m_A=2$ TeV and
$\mu$ variable over the range $\mu\sim 100-500$ GeV.
This model has gluinos and top-squarks with masses beyond LHC bounds and has
$m_h\simeq 125$ GeV with electroweak naturalness $\Delta_{EW}\sim 20-30$. 

There are actually 8 distinct higgsino pair production reactions:
$pp\to \tchi_1^\pm\tchi_{1,2}^0$, $\tchi_{1,2}^0\tchi_{1,2}^0$ and $\tchi_1^+\tchi_1^-$. In RPC SUSY, these reactions are somewhat hidden\cite{Baer:2011ec}
since the heavier higgsinos decay to lighter higgsinos plus soft visible
energy which usually falls below ATLAS/CMS triggering requirements.
The bulk of the reaction energy disappears in the form of the LSP
$\tchi_1^0$ rest mass. The summed higgsino pair production rate, labelled as {\it total}, ranges from $10^2-10^4$ fb over the natural range of $\mu$.
Thus, with $\sim 139$ fb$^{-1}$ of integrated luminosity from LHC Run 2,
we would expect $\sim 10^4-10^6$ higgsino pair events to lie within the
LHC Run 2 data sample.

Under RPV, then the higgsino pairs are no longer hidden.
Some of the various expected RPV event topologies are explored in
Ref's \cite{Dreiner:2023bvs} and \cite{Dreiner:2025kfd}.
Basically, we expect the following.
\bi
\item From dominant $\lambda_{ijk}$ couplings, higgsino pair production
  would result in $(\ell\bar{\ell}^\prime\nu_\ell )(\ell\bar{\ell}^\prime\nu_\ell )$
  final states which then contain four hard, isolated leptons plus $\eslt$.
  This class of events would be hard to miss at such high rates at
  LHC detectors.
\item From dominant $\lambda_{ijk}^\prime$ couplings, then we
  expect $(\ell q\bar{q}^\prime$ or $\nu q\bar{q}^\prime$)$^2$ events
  containing several hard jets, occasional isolated leptons and $\eslt$.
  A distinctive feature of such events would be that two of the jets plus an
  isolated lepton would reconstruct a resonance at $m_{\tchi_1^0}$.
\item From dominant $\lambda_{ijk}^{\prime\prime}$ couplings,
  then we get $(qq^\prime q^{\prime\prime})(qq^\prime q^{\prime\prime})$ events:
  hard multijet (possibly containing heavy flavors) wherein three jets
  on one side plus the other three jets would resonantly reconstruct
  the $m_{\tchi_1^0}$.
\ei

There have been some searches by ATLAS\cite{Ducu:2024xbx} and CMS\cite{CMS:2024ldy} for such topologies,
but it is unclear how well the simplified models which are used
correspond to the realistic situation arising from natural SUSY
with unstable LSPs. But as more data accrues in Run 3 and beyond,
the non-emergence of the (very clear) RPV signals would mean
rough upper limits on the various RPV couplings so as to render the higgsino
pair production reactions quasi-visible and where the LSP escapes detection
(corresponding to the white unshaded region of Fig. \ref{fig:tau_z1}.)

\section{Conclusions}
\label{sec:conclude}

In this paper, we have explored several aspects from a new theory of SUSY
dark matter set forth in Ref.~\cite{Baer:2025oid}.
This theory starts by assuming the MSSM supplemented with certain
anomaly-free discrete $R$-symmetries $\mathbb{Z}_n^R$
(for $n=4,6,8,12$ or 24) with matter $R$-charges consistent with
the existence of SUSY GUT multiplets. The $\mathbb{Z}_n^R$ symmetries
suppress the $\mu$ term (thus providing the first step in solving the
SUSY $\mu$ problem), and also forbid the various superpotential RPV couplings and dim-5 $p$-decay operators whilst allowing for the needed MSSM Yukawa terms
and see-saw neutrino terms. By augmenting the MSSM with
gauge singlet but $R$-charged  fields $X$ and $Y$, then under SUSY breaking the
$X$ and $Y$ fields develop intermediate scale VEVs leading to generation of the $\mu$ term ala Kim-Nilles. The theory admits an accidental global $U(1)_{PQ}$
which is broken as a consequence of SUSY breaking, and the corresponding Goldstone boson functions as a SUSY DFSZ axion, thus solving the strong CP problem.

Of note here is that the expected $R$-parity conservation can be also broken
via higher-dimensional operators leading  to RPC, like PQ, emerging as
an accidental, approximate symmetry.
In Sec. \ref{sec:basemodels}, we catalogued
the various possible models along with the relevant operator suppression.
In some cases, $R$-parity is conserved, or else conserved with sufficient
quality as to render the LSP stable on the timescale of the age of the universe.
In other cases, the RPV operators are suppressed by $\sim f_a/m_P$ leading
to the scenario where both WIMPs and SUSY DFSZ axions are produced in the
early universe, but the WIMPs decay away during the time between neutralino
freeze-out and the onset of BBN. Then only the DFSZ axions are left to
comprise all the SUSY dark matter. This scenario seems consistent with
recent negative WIMP direct detection search results from the LZ experiment,
where the search results are approaching the so-called neutrino floor.
With RPV couplings of order $f_a/m_P$, then some additional suppression,
perhaps in the form of discrete gauge symmetries, will be needed in order
to obey RPV $p$-decay constraints.

Along with these results, in Sec. \ref{sec:LSPdecay} we firmed up calculations of
the LSP decay rate in RPV models via exact tree-level results using
MadGraph including RPV decays.
The exact results contain all gaugino and higgsino mixing angles
and Yukawa terms in the couplings along with all diagrams with varying intermediate sfermion masses and phase space effects.
These results are in rough accord with simplistic analytic expressions,
validating earlier order-of-magnitude estimates.
We also commented in Sec. \ref{sec:lhc} that assuming that nature is natural,
with light higgsinos in the 100-350 GeV regime, then prompt LSP decays
should leave very compelling signatures that likely should have already been observed. The fact that no signals so far have been seen supports the case that
any RPV couplings are small, $\alt 10^{-4}-10^{-2}$, in which case the
LSPs would exit the detector, rendering the higgsino pair signals only quasi-visible with very soft visible energy release that can be difficult to
trigger on.

{\it Acknowledgements:} 
We thank X. Tata for comments on the manuscript.
HB gratefully acknowledges support from the Avenir Foundation.
VB gratefully acknowledges support from the William F. Vilas estate.

%%%%%%%%%%%%%%%%%%%%%%%%%%%%%%%%%%%%%%%%%%%%%%%%%%%%%%

%\section*{References}
\bibliography{natrpv}

\begin{thebibliography}{10}
\expandafter\ifx\csname url\endcsname\relax
  \def\url#1{\texttt{#1}}\fi
\expandafter\ifx\csname urlprefix\endcsname\relax\def\urlprefix{URL }\fi
\expandafter\ifx\csname href\endcsname\relax
  \def\href#1#2{#2} \def\path#1{#1}\fi

\bibitem{Martin:1997ns}
S.~P. Martin, {A Supersymmetry primer}, Adv. Ser. Direct. High Energy Phys. 18
  (1998) 1--98.
\newblock \href {http://arxiv.org/abs/hep-ph/9709356}
  {\path{arXiv:hep-ph/9709356}}, \href
  {https://doi.org/10.1142/9789812839657_0001}
  {\path{doi:10.1142/9789812839657_0001}}.

\bibitem{Baer:2006rs}
H.~Baer, X.~Tata, {Weak scale supersymmetry: From superfields to scattering
  events}, Cambridge University Press, 2006.

\bibitem{Bae:2019dgg}
K.~J. Bae, H.~Baer, V.~Barger, D.~Sengupta, {Revisiting the SUSY $\mu$ problem
  and its solutions in the LHC era}, Phys. Rev. D 99~(11) (2019) 115027.
\newblock \href {http://arxiv.org/abs/1902.10748} {\path{arXiv:1902.10748}},
  \href {https://doi.org/10.1103/PhysRevD.99.115027}
  {\path{doi:10.1103/PhysRevD.99.115027}}.

\bibitem{Giudice:1988yz}
G.~F. Giudice, A.~Masiero, {A Natural Solution to the mu Problem in
  Supergravity Theories}, Phys. Lett. B 206 (1988) 480--484.
\newblock \href {https://doi.org/10.1016/0370-2693(88)91613-9}
  {\path{doi:10.1016/0370-2693(88)91613-9}}.

\bibitem{Kim:1983dt}
J.~E. Kim, H.~P. Nilles, {The mu Problem and the Strong CP Problem}, Phys.
  Lett. B 138 (1984) 150--154.
\newblock \href {https://doi.org/10.1016/0370-2693(84)91890-2}
  {\path{doi:10.1016/0370-2693(84)91890-2}}.

\bibitem{Goity:1994dq}
J.~L. Goity, M.~Sher, {Bounds on $\Delta B = 1$ couplings in the supersymmetric
  standard model}, Phys. Lett. B 346 (1995) 69--74, [Erratum: Phys.Lett.B 385,
  500 (1996)].
\newblock \href {http://arxiv.org/abs/hep-ph/9412208}
  {\path{arXiv:hep-ph/9412208}}, \href
  {https://doi.org/10.1016/0370-2693(94)01688-9}
  {\path{doi:10.1016/0370-2693(94)01688-9}}.

\bibitem{Barbier:2004ez}
R.~Barbier, et~al., {R-parity violating supersymmetry}, Phys. Rept. 420 (2005)
  1--202.
\newblock \href {http://arxiv.org/abs/hep-ph/0406039}
  {\path{arXiv:hep-ph/0406039}}, \href
  {https://doi.org/10.1016/j.physrep.2005.08.006}
  {\path{doi:10.1016/j.physrep.2005.08.006}}.

\bibitem{Barger:1989rk}
V.~D. Barger, G.~F. Giudice, T.~Han, {Some New Aspects of Supersymmetry
  R-Parity Violating Interactions}, Phys. Rev. D 40 (1989) 2987.
\newblock \href {https://doi.org/10.1103/PhysRevD.40.2987}
  {\path{doi:10.1103/PhysRevD.40.2987}}.

\bibitem{Dreiner:1997uz}
H.~K. Dreiner, {An Introduction to explicit R-parity violation}, Adv. Ser.
  Direct. High Energy Phys. 21 (2010) 565--583.
\newblock \href {http://arxiv.org/abs/hep-ph/9707435}
  {\path{arXiv:hep-ph/9707435}}, \href
  {https://doi.org/10.1142/9789814307505_0017}
  {\path{doi:10.1142/9789814307505_0017}}.

\bibitem{Bhattacharyya:1997vv}
G.~Bhattacharyya, {A Brief review of R-parity violating couplings}, in:
  {Workshop on Physics Beyond the Standard Model: Beyond the Desert:
  Accelerator and Nonaccelerator Approaches}, 1997, pp. 194--201.
\newblock \href {http://arxiv.org/abs/hep-ph/9709395}
  {\path{arXiv:hep-ph/9709395}}.

\bibitem{Allanach:1999ic}
B.~C. Allanach, A.~Dedes, H.~K. Dreiner, {Bounds on R-parity violating
  couplings at the weak scale and at the GUT scale}, Phys. Rev. D 60 (1999)
  075014.
\newblock \href {http://arxiv.org/abs/hep-ph/9906209}
  {\path{arXiv:hep-ph/9906209}}, \href
  {https://doi.org/10.1103/PhysRevD.60.075014}
  {\path{doi:10.1103/PhysRevD.60.075014}}.

\bibitem{Hinchliffe:1992ad}
I.~Hinchliffe, T.~Kaeding, {B+L violating couplings in the minimal
  supersymmetric Standard Model}, Phys. Rev. D 47 (1993) 279--284.
\newblock \href {https://doi.org/10.1103/PhysRevD.47.279}
  {\path{doi:10.1103/PhysRevD.47.279}}.

\bibitem{Krauss:1988zc}
L.~M. Krauss, F.~Wilczek, {Discrete Gauge Symmetry in Continuum Theories},
  Phys. Rev. Lett. 62 (1989) 1221.
\newblock \href {https://doi.org/10.1103/PhysRevLett.62.1221}
  {\path{doi:10.1103/PhysRevLett.62.1221}}.

\bibitem{Ibanez:1991pr}
L.~E. Ibanez, G.~G. Ross, {Discrete gauge symmetries and the origin of baryon
  and lepton number conservation in supersymmetric versions of the standard
  model}, Nucl. Phys. B 368 (1992) 3--37.
\newblock \href {https://doi.org/10.1016/0550-3213(92)90195-H}
  {\path{doi:10.1016/0550-3213(92)90195-H}}.

\bibitem{Ibanez:1991hv}
L.~E. Ibanez, G.~G. Ross, {Discrete gauge symmetry anomalies}, Phys. Lett. B
  260 (1991) 291--295.
\newblock \href {https://doi.org/10.1016/0370-2693(91)91614-2}
  {\path{doi:10.1016/0370-2693(91)91614-2}}.

\bibitem{Buchmuller:2005sh}
W.~Buchmuller, K.~Hamaguchi, O.~Lebedev, M.~Ratz, {Local grand unification},
  in: {GUSTAVOFEST: Symposium in Honor of Gustavo C. Branco: CP Violation and
  the Flavor Puzzle}, 2005, pp. 143--156.
\newblock \href {http://arxiv.org/abs/hep-ph/0512326}
  {\path{arXiv:hep-ph/0512326}}.

\bibitem{Nilles:2009yd}
H.~P. Nilles, S.~Ramos-Sanchez, P.~K.~S. Vaudrevange, {Local Grand Unification
  and String Theory}, AIP Conf. Proc. 1200~(1) (2010) 226--234.
\newblock \href {http://arxiv.org/abs/0909.3948} {\path{arXiv:0909.3948}},
  \href {https://doi.org/10.1063/1.3327561} {\path{doi:10.1063/1.3327561}}.

\bibitem{Dreiner:2005rd}
H.~K. Dreiner, C.~Luhn, M.~Thormeier, {What is the discrete gauge symmetry of
  the MSSM?}, Phys. Rev. D 73 (2006) 075007.
\newblock \href {http://arxiv.org/abs/hep-ph/0512163}
  {\path{arXiv:hep-ph/0512163}}, \href
  {https://doi.org/10.1103/PhysRevD.73.075007}
  {\path{doi:10.1103/PhysRevD.73.075007}}.

\bibitem{Chen:2012tia}
M.-C. Chen, M.~Fallbacher, M.~Ratz, {Supersymmetric unification and R
  symmetries}, Mod. Phys. Lett. A 27 (2012) 1230044.
\newblock \href {http://arxiv.org/abs/1211.6247} {\path{arXiv:1211.6247}},
  \href {https://doi.org/10.1142/S0217732312300443}
  {\path{doi:10.1142/S0217732312300443}}.

\bibitem{Lee:2011dya}
H.~M. Lee, S.~Raby, M.~Ratz, G.~G. Ross, R.~Schieren, K.~Schmidt-Hoberg,
  P.~K.~S. Vaudrevange, {Discrete R symmetries for the MSSM and its singlet
  extensions}, Nucl. Phys. B 850 (2011) 1--30.
\newblock \href {http://arxiv.org/abs/1102.3595} {\path{arXiv:1102.3595}},
  \href {https://doi.org/10.1016/j.nuclphysb.2011.04.009}
  {\path{doi:10.1016/j.nuclphysb.2011.04.009}}.

\bibitem{Nilles:2012cy}
H.~P. Nilles, M.~Ratz, P.~K.~S. Vaudrevange, {Origin of Family Symmetries},
  Fortsch. Phys. 61 (2013) 493--506.
\newblock \href {http://arxiv.org/abs/1204.2206} {\path{arXiv:1204.2206}},
  \href {https://doi.org/10.1002/prop.201200120}
  {\path{doi:10.1002/prop.201200120}}.

\bibitem{Bhattiprolu:2021rrj}
P.~N. Bhattiprolu, S.~P. Martin, {High-quality axions in solutions to the
  \ensuremath{\mu} problem}, Phys. Rev. D 104~(5) (2021) 055014.
\newblock \href {http://arxiv.org/abs/2106.14964} {\path{arXiv:2106.14964}},
  \href {https://doi.org/10.1103/PhysRevD.104.055014}
  {\path{doi:10.1103/PhysRevD.104.055014}}.

\bibitem{Murayama:1992dj}
H.~Murayama, H.~Suzuki, T.~Yanagida, {Radiative breaking of Peccei-Quinn
  symmetry at the intermediate mass scale}, Phys. Lett. B 291 (1992) 418--425.
\newblock \href {https://doi.org/10.1016/0370-2693(92)91397-R}
  {\path{doi:10.1016/0370-2693(92)91397-R}}.

\bibitem{Baer:2018avn}
H.~Baer, V.~Barger, D.~Sengupta, {Gravity safe, electroweak natural axionic
  solution to strong $CP$ and SUSY $\mu$ problems}, Phys. Lett. B 790 (2019)
  58--63.
\newblock \href {http://arxiv.org/abs/1810.03713} {\path{arXiv:1810.03713}},
  \href {https://doi.org/10.1016/j.physletb.2019.01.007}
  {\path{doi:10.1016/j.physletb.2019.01.007}}.

\bibitem{Choi:1996vz}
K.~Choi, E.~J. Chun, J.~E. Kim, {Cosmological implications of radiatively
  generated axion scale}, Phys. Lett. B 403 (1997) 209--217.
\newblock \href {http://arxiv.org/abs/hep-ph/9608222}
  {\path{arXiv:hep-ph/9608222}}, \href
  {https://doi.org/10.1016/S0370-2693(97)00465-6}
  {\path{doi:10.1016/S0370-2693(97)00465-6}}.

\bibitem{Martin:2000eq}
S.~P. Martin, {Collider signals from slow decays in supersymmetric models with
  an intermediate scale solution to the mu problem}, Phys. Rev. D 62 (2000)
  095008.
\newblock \href {http://arxiv.org/abs/hep-ph/0005116}
  {\path{arXiv:hep-ph/0005116}}, \href
  {https://doi.org/10.1103/PhysRevD.62.095008}
  {\path{doi:10.1103/PhysRevD.62.095008}}.

\bibitem{Babu:2002ic}
K.~S. Babu, I.~Gogoladze, K.~Wang, {Stabilizing the axion by discrete gauge
  symmetries}, Phys. Lett. B 560 (2003) 214--222.
\newblock \href {http://arxiv.org/abs/hep-ph/0212339}
  {\path{arXiv:hep-ph/0212339}}, \href
  {https://doi.org/10.1016/S0370-2693(03)00411-8}
  {\path{doi:10.1016/S0370-2693(03)00411-8}}.

\bibitem{Lee:2010gv}
H.~M. Lee, S.~Raby, M.~Ratz, G.~G. Ross, R.~Schieren, K.~Schmidt-Hoberg,
  P.~K.~S. Vaudrevange, {A unique $\mathbb{Z}_4^R$ symmetry for the MSSM},
  Phys. Lett. B 694 (2011) 491--495.
\newblock \href {http://arxiv.org/abs/1009.0905} {\path{arXiv:1009.0905}},
  \href {https://doi.org/10.1016/j.physletb.2010.10.038}
  {\path{doi:10.1016/j.physletb.2010.10.038}}.

\bibitem{Bae:2013bva}
K.~J. Bae, H.~Baer, E.~J. Chun, {Mainly axion cold dark matter from natural
  supersymmetry}, Phys. Rev. D 89~(3) (2014) 031701.
\newblock \href {http://arxiv.org/abs/1309.0519} {\path{arXiv:1309.0519}},
  \href {https://doi.org/10.1103/PhysRevD.89.031701}
  {\path{doi:10.1103/PhysRevD.89.031701}}.

\bibitem{Bae:2013hma}
K.~J. Bae, H.~Baer, E.~J. Chun, {Mixed axion/neutralino dark matter in the SUSY
  DFSZ axion model}, JCAP 12 (2013) 028.
\newblock \href {http://arxiv.org/abs/1309.5365} {\path{arXiv:1309.5365}},
  \href {https://doi.org/10.1088/1475-7516/2013/12/028}
  {\path{doi:10.1088/1475-7516/2013/12/028}}.

\bibitem{Bae:2014rfa}
K.~J. Bae, H.~Baer, A.~Lessa, H.~Serce, {Coupled Boltzmann computation of mixed
  axion neutralino dark matter in the SUSY DFSZ axion model}, JCAP 10 (2014)
  082.
\newblock \href {http://arxiv.org/abs/1406.4138} {\path{arXiv:1406.4138}},
  \href {https://doi.org/10.1088/1475-7516/2014/10/082}
  {\path{doi:10.1088/1475-7516/2014/10/082}}.

\bibitem{Bae:2017hlp}
K.~J. Bae, H.~Baer, H.~Serce, {Prospects for axion detection in natural SUSY
  with mixed axion-higgsino dark matter: back to invisible?}, JCAP 06 (2017)
  024.
\newblock \href {http://arxiv.org/abs/1705.01134} {\path{arXiv:1705.01134}},
  \href {https://doi.org/10.1088/1475-7516/2017/06/024}
  {\path{doi:10.1088/1475-7516/2017/06/024}}.

\bibitem{Baer:2025oid}
H.~Baer, V.~Barger, D.~Sengupta, K.~Zhang, {All axion dark matter from
  supersymmetric models} (2 2025).
\newblock \href {http://arxiv.org/abs/2502.06955} {\path{arXiv:2502.06955}}.

\bibitem{Baer:2012up}
H.~Baer, V.~Barger, P.~Huang, A.~Mustafayev, X.~Tata, {Radiative natural SUSY
  with a 125 GeV Higgs boson}, Phys. Rev. Lett. 109 (2012) 161802.
\newblock \href {http://arxiv.org/abs/1207.3343} {\path{arXiv:1207.3343}},
  \href {https://doi.org/10.1103/PhysRevLett.109.161802}
  {\path{doi:10.1103/PhysRevLett.109.161802}}.

\bibitem{Dine:1982ah}
M.~Dine, W.~Fischler, {The Not So Harmless Axion}, Phys. Lett. B 120 (1983)
  137--141.
\newblock \href {https://doi.org/10.1016/0370-2693(83)90639-1}
  {\path{doi:10.1016/0370-2693(83)90639-1}}.

\bibitem{Abbott:1982af}
L.~F. Abbott, P.~Sikivie, {A Cosmological Bound on the Invisible Axion}, Phys.
  Lett. B 120 (1983) 133--136.
\newblock \href {https://doi.org/10.1016/0370-2693(83)90638-X}
  {\path{doi:10.1016/0370-2693(83)90638-X}}.

\bibitem{Preskill:1982cy}
J.~Preskill, M.~B. Wise, F.~Wilczek, {Cosmology of the Invisible Axion}, Phys.
  Lett. B 120 (1983) 127--132.
\newblock \href {https://doi.org/10.1016/0370-2693(83)90637-8}
  {\path{doi:10.1016/0370-2693(83)90637-8}}.

\bibitem{LZ:2024zvo}
J.~Aalbers, et~al., {Dark Matter Search Results from 4.2 Tonne-Years of
  Exposure of the LUX-ZEPLIN (LZ) Experiment} (10 2024).
\newblock \href {http://arxiv.org/abs/2410.17036} {\path{arXiv:2410.17036}}.

\bibitem{Baer:2013vpa}
H.~Baer, V.~Barger, D.~Mickelson, {Direct and indirect detection of
  higgsino-like WIMPs: concluding the story of electroweak naturalness}, Phys.
  Lett. B 726 (2013) 330--336.
\newblock \href {http://arxiv.org/abs/1303.3816} {\path{arXiv:1303.3816}},
  \href {https://doi.org/10.1016/j.physletb.2013.08.060}
  {\path{doi:10.1016/j.physletb.2013.08.060}}.

\bibitem{Baer:2016ucr}
H.~Baer, V.~Barger, H.~Serce, {SUSY under siege from direct and indirect WIMP
  detection experiments}, Phys. Rev. D 94~(11) (2016) 115019.
\newblock \href {http://arxiv.org/abs/1609.06735} {\path{arXiv:1609.06735}},
  \href {https://doi.org/10.1103/PhysRevD.94.115019}
  {\path{doi:10.1103/PhysRevD.94.115019}}.

\bibitem{Goldberg:1983nd}
H.~Goldberg, {Constraint on the Photino Mass from Cosmology}, Phys. Rev. Lett.
  50 (1983) 1419, [Erratum: Phys.Rev.Lett. 103, 099905 (2009)].
\newblock \href {https://doi.org/10.1103/PhysRevLett.50.1419}
  {\path{doi:10.1103/PhysRevLett.50.1419}}.

\bibitem{Ellis:1983ew}
J.~R. Ellis, J.~S. Hagelin, D.~V. Nanopoulos, K.~A. Olive, M.~Srednicki,
  {Supersymmetric Relics from the Big Bang}, Nucl. Phys. B 238 (1984) 453--476.
\newblock \href {https://doi.org/10.1016/0550-3213(84)90461-9}
  {\path{doi:10.1016/0550-3213(84)90461-9}}.

\bibitem{Jungman:1995df}
G.~Jungman, M.~Kamionkowski, K.~Griest, {Supersymmetric dark matter}, Phys.
  Rept. 267 (1996) 195--373.
\newblock \href {http://arxiv.org/abs/hep-ph/9506380}
  {\path{arXiv:hep-ph/9506380}}, \href
  {https://doi.org/10.1016/0370-1573(95)00058-5}
  {\path{doi:10.1016/0370-1573(95)00058-5}}.

\bibitem{Baer:2011ec}
H.~Baer, V.~Barger, P.~Huang, {Hidden SUSY at the LHC: the light higgsino-world
  scenario and the role of a lepton collider}, JHEP 11 (2011) 031.
\newblock \href {http://arxiv.org/abs/1107.5581} {\path{arXiv:1107.5581}},
  \href {https://doi.org/10.1007/JHEP11(2011)031}
  {\path{doi:10.1007/JHEP11(2011)031}}.

\bibitem{Bae:2014yta}
K.~J. Bae, H.~Baer, H.~Serce, {Natural little hierarchy for SUSY from radiative
  breaking of the Peccei-Quinn symmetry}, Phys. Rev. D 91~(1) (2015) 015003.
\newblock \href {http://arxiv.org/abs/1410.7500} {\path{arXiv:1410.7500}},
  \href {https://doi.org/10.1103/PhysRevD.91.015003}
  {\path{doi:10.1103/PhysRevD.91.015003}}.

\bibitem{Bagger:1993ji}
J.~Bagger, E.~Poppitz, {Destabilizing divergences in supergravity coupled
  supersymmetric theories}, Phys. Rev. Lett. 71 (1993) 2380--2382.
\newblock \href {http://arxiv.org/abs/hep-ph/9307317}
  {\path{arXiv:hep-ph/9307317}}, \href
  {https://doi.org/10.1103/PhysRevLett.71.2380}
  {\path{doi:10.1103/PhysRevLett.71.2380}}.

\bibitem{Baer:2017uvn}
H.~Baer, V.~Barger, H.~Serce, K.~Sinha, {Higgs and superparticle mass
  predictions from the landscape}, JHEP 03 (2018) 002.
\newblock \href {http://arxiv.org/abs/1712.01399} {\path{arXiv:1712.01399}},
  \href {https://doi.org/10.1007/JHEP03(2018)002}
  {\path{doi:10.1007/JHEP03(2018)002}}.

\bibitem{Paige:2003mg}
F.~E. Paige, S.~D. Protopopescu, H.~Baer, X.~Tata, {ISAJET 7.69: A Monte Carlo
  event generator for pp, anti-p p, and e+e- reactions} (12 2003).
\newblock \href {http://arxiv.org/abs/hep-ph/0312045}
  {\path{arXiv:hep-ph/0312045}}.

\bibitem{Skands:2003cj}
P.~Z. Skands, et~al., {SUSY Les Houches accord: Interfacing SUSY spectrum
  calculators, decay packages, and event generators}, JHEP 07 (2004) 036.
\newblock \href {http://arxiv.org/abs/hep-ph/0311123}
  {\path{arXiv:hep-ph/0311123}}, \href
  {https://doi.org/10.1088/1126-6708/2004/07/036}
  {\path{doi:10.1088/1126-6708/2004/07/036}}.

\bibitem{Alwall:2011uj}
J.~Alwall, M.~Herquet, F.~Maltoni, O.~Mattelaer, T.~Stelzer, {MadGraph 5 :
  Going Beyond}, JHEP 06 (2011) 128.
\newblock \href {http://arxiv.org/abs/1106.0522} {\path{arXiv:1106.0522}},
  \href {https://doi.org/10.1007/JHEP06(2011)128}
  {\path{doi:10.1007/JHEP06(2011)128}}.

\bibitem{Alwall:2014hca}
J.~Alwall, R.~Frederix, S.~Frixione, V.~Hirschi, F.~Maltoni, O.~Mattelaer,
  H.~S. Shao, T.~Stelzer, P.~Torrielli, M.~Zaro, {The automated computation of
  tree-level and next-to-leading order differential cross sections, and their
  matching to parton shower simulations}, JHEP 07 (2014) 079.
\newblock \href {http://arxiv.org/abs/1405.0301} {\path{arXiv:1405.0301}},
  \href {https://doi.org/10.1007/JHEP07(2014)079}
  {\path{doi:10.1007/JHEP07(2014)079}}.

\bibitem{RPVUFO}
\url{https://github.com/ilmonteux/RPVMSSM_UFO} (2019).

\bibitem{Ellis:2002wv}
J.~R. Ellis, K.~A. Olive, Y.~Santoso, {The MSSM parameter space with
  nonuniversal Higgs masses}, Phys. Lett. B 539 (2002) 107--118.
\newblock \href {http://arxiv.org/abs/hep-ph/0204192}
  {\path{arXiv:hep-ph/0204192}}, \href
  {https://doi.org/10.1016/S0370-2693(02)02071-3}
  {\path{doi:10.1016/S0370-2693(02)02071-3}}.

\bibitem{Jedamzik:2006xz}
K.~Jedamzik, {Big bang nucleosynthesis constraints on hadronically and
  electromagnetically decaying relic neutral particles}, Phys. Rev. D 74 (2006)
  103509.
\newblock \href {http://arxiv.org/abs/hep-ph/0604251}
  {\path{arXiv:hep-ph/0604251}}, \href
  {https://doi.org/10.1103/PhysRevD.74.103509}
  {\path{doi:10.1103/PhysRevD.74.103509}}.

\bibitem{Beenakker:1996ed}
W.~Beenakker, R.~Hopker, M.~Spira, {PROSPINO: A Program for the production of
  supersymmetric particles in next-to-leading order QCD} (11 1996).
\newblock \href {http://arxiv.org/abs/hep-ph/9611232}
  {\path{arXiv:hep-ph/9611232}}.

\bibitem{Ellis:2002iu}
J.~R. Ellis, T.~Falk, K.~A. Olive, Y.~Santoso, {Exploration of the MSSM with
  nonuniversal Higgs masses}, Nucl. Phys. B 652 (2003) 259--347.
\newblock \href {http://arxiv.org/abs/hep-ph/0210205}
  {\path{arXiv:hep-ph/0210205}}, \href
  {https://doi.org/10.1016/S0550-3213(02)01144-6}
  {\path{doi:10.1016/S0550-3213(02)01144-6}}.

\bibitem{Baer:2005bu}
H.~Baer, A.~Mustafayev, S.~Profumo, A.~Belyaev, X.~Tata, {Direct, indirect and
  collider detection of neutralino dark matter in SUSY models with
  non-universal Higgs masses}, JHEP 07 (2005) 065.
\newblock \href {http://arxiv.org/abs/hep-ph/0504001}
  {\path{arXiv:hep-ph/0504001}}, \href
  {https://doi.org/10.1088/1126-6708/2005/07/065}
  {\path{doi:10.1088/1126-6708/2005/07/065}}.

\bibitem{Dreiner:2023bvs}
H.~K. Dreiner, Y.~S. Koay, D.~K\"ohler, V.~M. Lozano, J.~Montejo~Berlingen,
  S.~Nangia, N.~Strobbe, {The ABC of RPV: classification of R-parity violating
  signatures at the LHC for small couplings}, JHEP 07 (2023) 215.
\newblock \href {http://arxiv.org/abs/2306.07317} {\path{arXiv:2306.07317}},
  \href {https://doi.org/10.1007/JHEP07(2023)215}
  {\path{doi:10.1007/JHEP07(2023)215}}.

\bibitem{Dreiner:2025kfd}
H.~K. Dreiner, M.~Hank, Y.~S. Koay, M.~Sch\"urmann, R.~Sengupta, A.~Shah,
  N.~Strobbe, E.~Thomson, {The ABC of RPV II: Classification of R-parity
  Violating Signatures from UDD Couplings and their Coverage at the LHC} (3
  2025).
\newblock \href {http://arxiv.org/abs/2503.03830} {\path{arXiv:2503.03830}}.

\bibitem{Ducu:2024xbx}
O.~Ducu, {ATLAS searches for higgsinos with R-parity violating couplings in
  events with leptons}, Int. J. Mod. Phys. A 40~(08) (2025) 2443014.
\newblock \href {http://arxiv.org/abs/2412.19317} {\path{arXiv:2412.19317}},
  \href {https://doi.org/10.1142/S0217751X24430140}
  {\path{doi:10.1142/S0217751X24430140}}.

\bibitem{CMS:2024ldy}
A.~Hayrapetyan, et~al., {Searches for Pair-Produced Multijet Resonances Using
  Data Scouting in Proton-Proton Collisions at s=13\,\,TeV}, Phys. Rev. Lett.
  133~(20) (2024) 201803.
\newblock \href {http://arxiv.org/abs/2404.02992} {\path{arXiv:2404.02992}},
  \href {https://doi.org/10.1103/PhysRevLett.133.201803}
  {\path{doi:10.1103/PhysRevLett.133.201803}}.

\end{thebibliography}
\bibliographystyle{elsarticle-num}

\end{document}